\documentclass[trackchanges,twocolumn,resetfootnote]{aastex701}

\usepackage{url}

\shorttitle{XRISM constraints on the 3.5 keV line}
\shortauthors{XRISM Collaboration}

\begin{document}

\title{XRISM constraints on unidentified X-ray emission lines, including the 3.5 keV line, in the stacked spectrum of ten galaxy clusters}


\correspondingauthor{Ming Sun (ms0071@uah.edu), Michael Loewenstein (michael.loewenstein-1@nasa.gov) and Maxim Markevitch (maxim.markevitch@nasa.gov)}

\collaboration{0}{XRISM Collaboration:%
}

\author[0000-0003-4721-034X]{Marc Audard}
\affiliation{Department of Astronomy, University of Geneva, Versoix CH-1290, Switzerland} 
\email{Marc.Audard@unige.ch}

\author{Hisamitsu Awaki}
\affiliation{Department of Physics, Ehime University, Ehime 790-8577, Japan}
\email{awaki@astro.phys.sci.ehime-u.ac.jp}

\author[0000-0002-1118-8470]{Ralf Ballhausen}
\affiliation{Department of Astronomy, University of Maryland, College Park, MD 20742, USA}
\affiliation{NASA / Goddard Space Flight Center, Greenbelt, MD 20771, USA}
\affiliation{Center for Research and Exploration in Space Science and Technology, NASA / GSFC (CRESST II), Greenbelt, MD 20771, USA}
\email{ballhaus@umd.edu}

\author[0000-0003-0890-4920]{Aya Bamba}
\affiliation{Department of Physics, University of Tokyo, Tokyo 113-0033, Japan}
\email{bamba@phys.s.u-tokyo.ac.jp}

\author[0000-0001-9735-4873]{Ehud Behar}
\affiliation{Department of Physics, Technion, Technion City, Haifa 3200003, Israel}
\email{behar@physics.technion.ac.il}

\author[0000-0003-2704-599X]{Rozenn Boissay-Malaquin}
\affiliation{Center for Space Sciences and Technology, University of Maryland, Baltimore County (UMBC), Baltimore, MD, 21250 USA}
\affiliation{NASA / Goddard Space Flight Center, Greenbelt, MD 20771, USA}
\affiliation{Center for Research and Exploration in Space Science and Technology, NASA / GSFC (CRESST II), Greenbelt, MD 20771, USA}
\email{rozennbm@umbc.edu}

\author[0000-0003-2663-1954]{Laura Brenneman}
\affiliation{Center for Astrophysics | Harvard-Smithsonian, Cambridge, MA 02138, USA}
\email{lbrenneman@cfa.harvard.edu}

\author[0000-0001-6338-9445]{Gregory V.\ Brown}
\affiliation{Lawrence Livermore National Laboratory, Livermore, CA 94550, USA}
\email{brown86@llnl.gov}

\author[0000-0002-5466-3817]{Lia Corrales}
\affiliation{Department of Astronomy, University of Michigan, Ann Arbor, MI 48109, USA}
\email{liac@umich.edu}

\author[0000-0001-8470-749X]{Elisa Costantini}
\affiliation{SRON Netherlands Institute for Space Research, Leiden, The Netherlands}
\email{e.costantini@sron.nl}

% ...

\begin{abstract}
We stack 3.75 Megaseconds of early XRISM Resolve observations of ten galaxy clusters to search for unidentified spectral lines in the $E=$ 2.5-15 keV band (rest frame), including the $E=3.5$ keV line reported in earlier, low spectral resolution studies of cluster samples. Such an emission line may originate from the decay of the sterile neutrino, a warm dark matter (DM) candidate. No unidentified lines are detected in our stacked cluster spectrum, with the $3\sigma$ upper limit on the $m_{\rm s}\sim$ 7.1 keV DM particle decay rate (which corresponds to a $E=3.55$ keV emission line) of $\Gamma \sim 1.0 \times 10^{-27}$ s$^{-1}$. This upper limit is $3-4$ times lower than the one derived by \citet{Markevitch17} from the Perseus observation, but still 5 times higher than the XMM-Newton detection reported by \citet{Bulbul14} in the stacked cluster sample. XRISM Resolve, with its high spectral resolution but a small field of view, may reach the sensitivity needed to test the XMM-Newton cluster sample detection by combining several years worth of future cluster observations.
\end{abstract}


\section{Introduction}
\label{sect:intro}

The search for decaying dark matter (DM) through X-ray emission lines has received considerable attention in recent years. Particularly, the reported detection of an unidentified emission line at 3.5 keV in the spectra of galaxy clusters, M31 and the Milky Way from XMM-Newton and Chandra (e.g., \citealt{Bulbul14}, hereafter B14; \citealt{boyarsky14,boyarsky15}; \citealt{cappelluti18}) has sparked intensive interest and followup studies, because such a line could arise from the radiative decay of a sterile neutrino with the mass $m_s\sim7$ keV --- a potential warm dark matter candidate \citep{dodelson94, abazajian17}. Sterile neutrinos are among the most extensively studied warm dark matter candidates. Their production mechanisms can naturally give rise to the observed present day dark matter density (the relic abundance) and, due to their warm nature, sterile neutrinos suppress the formation of small-scale structure. This helps alleviate long-standing small-scale structure challenges to cold dark matter, such as the missing satellites and core–cusp problems \citep[e.g.,][]{abazajian01,abazajian17,boyarsky19,2021PhR...928....1D}. The inferred fluxes and their corresponding mixing angles from the initial detections lie within the range of the viable warm dark matter models \citep{abazajian17}.

The XMM-Newton and Chandra detections were based on data with the modest spectral resolution of the CCD detectors ($100-120$ eV) and pushed the boundary of the technical capabilities of those instruments. The spectral stacking approach employed in B14, with the spectra of clusters at different redshifts coadded in the cluster reference frame, was designed to amplify any common cluster spectral features while diluting the contribution of any detector artifacts, such as the inaccuracies in the shape of the instrument effective area curves or detector background lines. Nevertheless, the detected signal was very faint and could still be affected by a number of modeling uncertainties, as described in detail in B14. These complications and uncertainties were subsequently discussed in \citep[e.g.,][]{jeltema15, 2015MNRAS.451.2447U, carlson15, Dessert24} and include modeling of the surrounding weak emission lines from the intracluster plasma, instrument calibration, and spectral fitting approaches. An alternative physical possibility for the line emission around the energy of the detection was also proposed --- the charge exchange between highly ionized sulfur in the hot plasma and cold gas in central cluster regions \citep{gu15,shah16}.

Other studies of the dark matter dominated systems such as the Milky Way and dwarf galaxies, using data from XMM-Newton, Chandra, NuSTAR and Swift, yielded non-detections of the 3.5 keV X-ray line at the level expected from the B14 cluster result under the assumption of its DM decay origin
\citep{malyshev14,anderson15,2016PhRvD..94l3504N,2017PhRvD..95l3002P,2019A&A...625L...7H,dessert20,sicilian20,roach20,foster21,sicilian22,roach23,Dessert24}, while a possible Chandra positive detection was reported for the cluster Zw3146 \citep{Bhargava24_Zw3146}.

The microcalorimeter onboard Hitomi was the first instrument to obtain high-resolution (5 eV) spectra of a galaxy cluster. It observed Perseus, the cluster for which B14 reported an anomalously bright 3.5 keV signal compared to the rest of their cluster sample. Hitomi did not detect the line in Perseus, ruling out the B14 Perseus signal at $>99$\% confidence \citep{Markevitch17, 2024PASJ...76..512F}, but it lacked the depth required to test the much lower line brightness based on the B14 stacked cluster data. It did uncover a hint of the sulfur charge exchange signal \citep{Hitomi18_atomic}. As pointed out in B14, a single-cluster detection can be affected by small instrumental artifacts much more strongly than a sample spanning a range of redshifts, which is the likely cause for the B14 Perseus detection. This also applies to multiple objects at the same $z=0$, such as the Milky Way and dwarf galaxies. It is therefore important to examine a large sample of clusters at different $z$\/ with a high-resolution instrument.

The recent launch of the Resolve instrument onboard XRISM, the successor to Hitomi \citep{Tashiro2020_XRISM, Ishisaki2022_Resolve}, provides such a capability (see e.g., \citealt{2023MNRAS.524.6345L,2024ApJ...976..238Z,2024PhRvL.132u1002D}).
An important thing to note is the small Resolve field of view (FOV, $3'\times 3'$), which makes its {\em grasp}\/ --- the product of the effective area and the solid angle covered, which is the quantity that determines the number of photons collected from an extended celestial source such as a nearby cluster --- much smaller than the grasp of XMM-Newton. Given the low brightness for the 3.5 keV line that have been discussed, Resolve would require a very long combined cluster exposure to approach the commensurate sensitivity. A Resolve study for the single Centaurus cluster was already reported in \citet{2025arXiv250304726Y}, where both the double-line
search and the single-line search has been done. While the limits are weaker or comparable than some existing ones, the work demonstrated the potential with more Resolve data.

In this work, we follow the B14 approach and stack the recent Resolve observations of 10 clusters at different $z$\/ (eight from the Performance Verification phase and two from the early General Observer program) in the source rest frame for a deep search for a possible DM line, taking advantage of the high energy resolution of the Resolve instrument ($\sim$ 5 eV full width at half-maximum, FWHM). This paper focuses on the 3.5 keV line but we also search for unidentified lines in the 2.5--15 keV interval probed by Resolve. 
Throughout this paper, we assume a $\Lambda$CDM cosmology with $H_0$ = 70 km s$^{-1}$ Mpc$^{-1}$, $\Omega_m$ = 0.3, and $\Omega_{\Lambda}$ = 0.7.
We use the solar abundance table from \cite{aspl09}.
Unless otherwise specified, uncertainties are $1\sigma$.

\begin{table*}[!t]
 \begin{center}
   \vspace{-0.1cm}
 \caption{The XRISM data on galaxy clusters for stacking}
   \label{tab:obs_log}
   \vspace{-0.4cm}
    \begin{tabular}{cccccc}\hline\hline
        Cluster & OBSID & Exp (ks)$^{a}$ & Counts$^{b}$ & $M_{\rm DM}$ (10$^{12}$ M$_{\odot}$)$^{c}$ & $w^{d}$ \\
        \hline
        Virgo (M87) ($z$=0.00428, $D_{\rm L}$=16.5 Mpc) & 300014010 & 116.8 & 2993 & 0.29 & 0.48 \\
        $M_{\rm 200c}=10^{14.02}$ M$_{\odot}$ \citep{Simionescu17}, (5.3, 36$'$)$^{\rm e}$ & 300015010 & 159.7 & 1659 & 0.20 &  \\
         & 300016010 & 169.7 & 1950 & 0.20 &  \\
            & 300016020 &  & &  &  \\
            & 300017010 & 76.1 & 789 & 0.19 &  \\
            & 300017020 & 82.0 & 820 & 0.19 &  \\
            \hline 
         Centaurus ($z$=0.01003, $D_{\rm L}$=36.8 Mpc)  & 000138000 & 287.4 & 4291 & 1.4 & 0.31 \\
         $M_{\rm 200c}=10^{14.36}$ M$_{\odot}$ \citep{2013MNRAS.432..554W}, (4.9, 23$'$) & & & & & \\
            \hline 
        Perseus ($z$=0.0179, $D_{\rm L}$=77.7 Mpc)    & 000154000 & 48.7 & 9605 & 6.9 & 0.58 \\
        $M_{\rm 200c}=10^{14.82}$ M$_{\odot}$ \citep{2011Sci...331.1576S}, (4.4, 18$'$)   & 000155000 & 53.3 & 10565 & 6.9 & \\
            & 000156000 & 58.5 & 3889 & 4.2 & \\
            & 000157000 & 99.0 & 1986 & 2.6 & \\
            & 000158000 & 133.3 & 1088 & 1.8 & \\
            & 101009010 & 46.8 & 10326 & 6.9 & \\
            & 101010010 & 42.8 & 8607 & 6.9 & \\
            & 101011010$^{f}$ & 40.0 & 7982 & 6.8 &  \\
            & 101012010 & 44.2 & 8888 & 6.9 &  \\
            & 201078010$^{g}$ & 54.2 & 4433 & 4.8 &  \\
            & 201080010 & 92.4 & 961 & 2.0 &  \\
            & 201079010 & 60.0 & 1030 & 2.9 & \\
            & 201079020 & 62.1 & 1110 & 2.9 & \\
            \hline
        Coma ($z$=0.0231, $D_{\rm L}$=101 Mpc)    & 300073010 & 397.6 & 4559 & 12 & 0.53 \\ 
        $M_{\rm 200c}=10^{14.95}$ M$_{\odot}$ \citep{PlanckComa}   & 300074010 & 158.4 & 1056 & 4.3 & \\
        (4.3, 16$'$)  & 300074020 &  & &  & \\
            \hline
        Ophiuchus ($z$=0.0296, $D_{\rm L}$= 130 Mpc)    & 201006010 & 217.1 & 14035 & 21 & 0.41 \\
        $M_{\rm 200c}=10^{15.20}$ M$_{\odot}$ \citep{2008PASJ...60.1133F}, (4.0, 15$'$)  & 201117010 & 116.6 & 5277 & 18 & \\
            \hline
        A2199 ($z$=0.0310, $D_{\rm L}$= 136 Mpc)    & 201089010 & 251.0 & 6559 & 12 & 0.16 \\
        $M_{\rm 200c}=10^{14.49}$ M$_{\odot}$ \citep{2020MNRAS.497.3943M}, (4.7, 7.4$'$) & & & & & \\
            \hline
        Hydra A ($z$=0.0543, $D_{\rm L}$=242 Mpc)    & 300073010 & 116.3 & 2389 & 25 & 0.051 \\
        $M_{\rm 200c}=10^{14.48}$ M$_{\odot}$ \citep{Ettori19}, (4.7, 4.3$'$) & & & & & \\
            \hline
        A2319 ($z$=0.0557, $D_{\rm L}$=249 Mpc)   & 000101000 & 55.8 & 941 & 37 & 0.14 \\
        $M_{\rm 200c}=10^{15.01}$ M$_{\odot}$ \citep{Ettori19}, (4.2, 7.1$'$)    & 000102000 & 49.1 & 1080 & 45 & \\
            & 000103000 & 88.8 & 1821 & 45 & \\
            \hline
        A2029 ($z$=0.0787, $D_{\rm L}$=357 Mpc)    & 000149000 & 12.4 & 495 & 79 & 0.11 \\
         $M_{\rm 200c}=10^{15.10}$ M$_{\odot}$ \citep{Ettori19}, (4.1, 5.6$'$)  & 000150000 & 102.2 & 851 & 36 & \\
            & 000151000 & 25.1 & 1021 & 79 & \\
            & 000152000 & 42.9 & 64 & 16  & \\
            & 300053010 & 366.2 & 526 & 16 & \\
        \hline
        PKS 0745-19 ($z$=0.103, $D_{\rm L}$=474 Mpc)    & 000149000$^{h}$ & 21.4 & 706 & 69 & 0.0072 \\
        $M_{\rm 200c}=10^{14.98}$ M$_{\odot}$ \citep{2012MNRAS.424.1826W}, (4.2, 3.9$'$) & & & & & \\
        \hline
    \end{tabular}
    \end{center}
    \vspace{-0.2cm}
Note: $^{\rm a}$ Net exposure time after data screening. The total exposure time used is 3.748 Ms. OBSID without exposure time is combined with the OBSID in the previous row as the same region is covered. $^{\rm b}$ Counts in the rest-frame $3.4-3.8$ keV, corrected with the relevant X-ray redshift. $^{\rm c}$ The projected DM mass within the XRISM Resolve FOV, with pixel 27 excluded. 
$^{\rm d}$ The $w$ factor for each cluster, is $(\sum_{i}{Exp_i * M_{\rm DM, i}}) * (1+z) / D_{\rm L}^2$, in units of 10$^{12}$ M$_{\odot}$ ksec Mpc$^{-2}$, where $i$ stands for each OBSID of the same cluster. The $w$ factor is proportional to the number of observed DM decay photons.
$^{e}$ The first number in the brackets is $c_{\rm 200c}$, while the second number is the scale radius of the DM NFW profile in arcmin. Same for all clusters in the table.
$^{f}$ Pixel 23 is also excluded. $^{g}$ Pixel 7 is also excluded. $^{h}$ 1/3 of the Resolve FOV excluded.
\end{table*}

\begin{figure*}[!t]
\begin{center}
\includegraphics[width=1.0\textwidth]{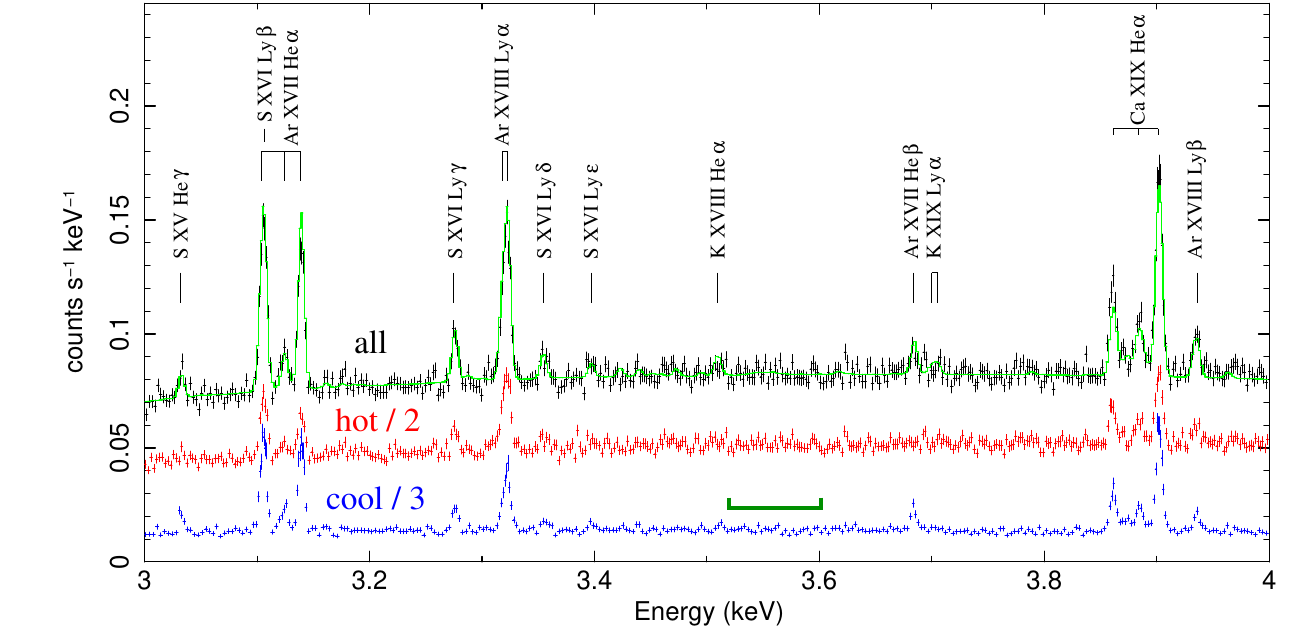}
\caption{The stacked XRISM spectra in the rest-frame 3--4 keV. Black shows the full cluster sample (ten clusters, 3.75 Ms total exposure), red shows the hot subsample (six clusters with $M_{\rm 200c} > 10^{14.5}$ M$_{\odot}$, 2.49 Ms total exposure), and blue for cool clusters in the sample (four clusters with $M_{\rm 200c} < 10^{14.5}$ M$_{\odot}$, 1.26 Ms total). The green curve shows the best fit with a two bapec model to the full sample.
For clarity, the hot cluster spectrum is lowered by a factor of two and the cool cluster spectrum is lowered by a factor of three.
Detected atomic lines in this energy range are marked. The atomic lines have a velocity dispersion of 150--160 km s$^{-1}$, which is $\sim$ 6 times smaller than the velocity dispersion we adopted for the DM line search in the full sample. For the cool and hot subsamples, the difference is $\sim$ 4 times and $\sim$ 7 times respectively. Thus, the expected DM line in these spectra should be 4--7 times broader than the shown atomic lines.
The green bracket shows the 90\% confidence interval on the unidentified 3.5 keV line energy for the most-restrictive XMM-Newton MOS stacked-clusters sample in B14.
}
\label{fig:spec}
\end{center}
\end{figure*}

\section{XRISM observations and data reduction}
\label{sect:obs}

The XRISM Resolve observations used in this work are summarized in Table~\ref{tab:obs_log}.
The detailed analysis of the XRISM Resolve data can be found in papers on individual clusters: Centaurus --- \cite{XRISMCenI}; A2029 --- \cite{XRISM2029a,XRISM2029b,XRISM2029c}; Coma --- \cite{XRISMcomaa}; Hydra~A --- \cite{Rose25}, Ophiuchus --- \cite{Fujita25}; A2319 --- \cite{XRISMA2319}; Perseus --- \cite{XRISMPers,2025arXiv251012782Z}.
A2319 and PKS\,0745--19 were observed early in the XRISM commissioning phase, which required a special calibration approach, discussed in detail in \cite{XRISMA2319}.
For each observation, the standard filters on the pulse invariant (PI), rise time, and pixel-pixel coincidence have been applied. Only high-resolution primary (Hp) events were included.
The Resolve FOV includes 35 0.5$'\times0.5'$ pixels.
Pixel 12, a continuously illuminated calibration pixel outside of the FOV \citep{2018JATIS...4a1214K}, and pixels exhibiting abrupt changes in their energy scales \citep{Porter2024} were excluded. Time-dependent changes in energy scale are corrected by interpolation between calibration measurements that use the 55Fe sources on the filter wheel, scheduled to map the slow variations associated with the recycling of the 50-mK cooler and with slewing, but this method cannot properly reconstruct the energy scale of pixels exhibiting abrupt changes in gain.  Pixel 27 exhibits frequent small gain jumps (several eV at 6 keV), and thus is always excluded. Several other pixels also experience gain jumps, but much less frequently. Those pixels were also excluded in the observations that included such a jump, as indicated in Table~\ref{tab:obs_log} and discussed in the corresponding papers.

For each observation, we extracted the spectrum for the full Resolve FOV from the cleaned event file. The instrument spectral resolution was modeled using the  redistribution matrix files (RMF) of the ``small'' size that includes only the Gaussian core of the line spread function, which is adequate for searching for faint lines. RMFs were generated for each observation using the respective event file filtered as described above, but with all pixels and grades included, while excluding the low-resolution secondary events and energies outside the studied range to minimize the presence of anomalous branching ratios.
These event files are used in this procedure to generate the relative count distribution across the detector array that determines the weights of the pixel-dependent line-spread-function implemented in the RMF generator.
For simplicity, on-axis point-source anciliary response files (ARF) were constructed by the \texttt{xaarfgen} ftool, using the spectral extraction detector region for each observation (Table~\ref{tab:obs_log}). This disregards the relatively small effect of the $1.3'$ (half-power diameter) telescope angular resolution on the predicted flux from an extended source that falls within the $3'$ FOV, which is adequate for our purpose.
Responses were generated using XRISM CalDB 11 (version 20250315), with adjustments in keeping with the de-redshifting of the individual spectra described below in \S~\ref{sect:stacking}.
The FOV-averaged energy scale uncertainty after the standard Resolve gain reconstruction is $\leq$0.3 eV in the 5.4--9 keV energy band \citep{Eckart2024, Porter2024}.
The energy scale uncertainty for $E<5.4$ keV becomes $\sim1$ eV because of the less precise calibration data
(Resolve team, private comm.), which is still much lower than the 5 eV Resolve energy resolution and the expected width of the DM line at 3.55 keV (11.3 eV for the full sample, as will be discussed below).

Most of the observations are pointed to the bright cluster centers and the contribution of the cosmic X-ray background (CXB) and the non-X-ray background (NXB) is small.
The CXB spectrum in our energy band (2--15 keV) is mainly the power law component from unresolved AGN and contributes $\sim$ 0.2\% of the 2.5--15 keV flux in the stacked spectrum, so we ignore it in our analysis.
The NXB spectrum is mostly a flat continuum with relatively bright narrow lines at $E=5.90$ keV, 7.47 keV, 9.71 keV and 11.44 keV (plus a few fainter lines)
\footnote{\url{https://heasarc.gsfc.nasa.gov/docs/xrism/analysis/nxb/nxb_spectral_models.html}}, and we generate a model of it 
for each of the 36 observations. The same redshift correction as applied to the cluster spectra (\S\ref{sect:stacking}) was also applied to each of those NXB models, which were then coadded to produce the NXB model for the stacked spectrum.
The NXB contribution in the 1.9--10 keV band is 1.2\% but increases to 6.3\% in the 9--10 keV band and 22\% in the 10--15 keV band.

\begin{figure*}[!t]
\begin{center}
\vspace{-0.4cm}
\includegraphics[width=0.96\textwidth]{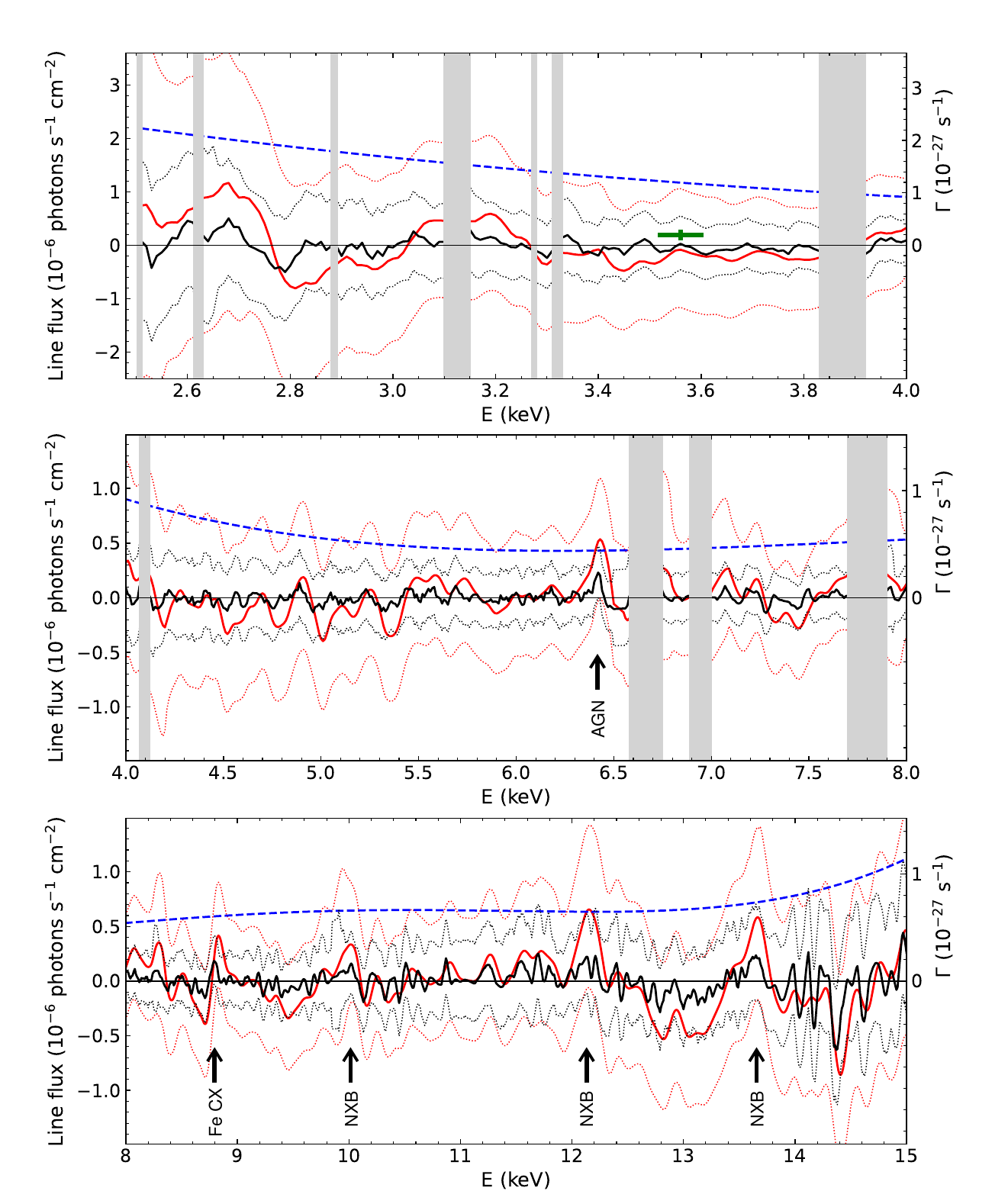}
\vspace{-0.1cm}
\caption{Best-fit flux for an additional line as a function of its energy across the 2.5--15 keV rest-frame band (split into three intervals for clarity). The red solid line shows the best-fit line flux for a line width of 950 km s$^{-1}$, the weighted velocity dispersion expected for DM in this sample, while the red dotted lines show the $3\sigma$ range. The black solid line shows the best-fit line flux for a width of 160 km s$^{-1}$ (an ICM line), while the black dotted lines show its $3\sigma$ range. The vertical grey shaded areas mark strong ICM emission lines, where the search for faint lines is not possible. 
The right vertical axis shows the approximate corresponding DM particle decay rate $\Gamma$, assuming all DM is comprised of the decaying particle.
The green cross shows the energy range and decay rate (and 1-$\sigma$ error) for the B14 3.5 keV line, 
which is still a factor 5 below our $3\sigma$ limit on the broad line in the same energy range.
The blue dashed line shows a polynomial fit to approximate the $3\sigma$ upper bound.
We also mark the positions of 6.40 keV Fe fluorescent line likely from X-ray AGN, the possible Fe charge exchange feature at $\sim$ 8.8 keV and three bumps at $E>$ 9.8 keV likely from residual NXB lines.
}
\label{fig:limits}
\end{center}
\end{figure*}

\section{Analysis and discussion}
\label{sect:analysis}

\subsection{Cluster mass modeling}
\label{sect:mass_model}

Since our goal is to constrain the signal from the DM decay, we need to estimate the projected DM mass within the Resolve FOV for each cluster and then the predicted signal from a stacked spectrum. A suitable mass model for each
cluster was identified from the literature,
with the values of $M_{\rm 200c}$ listed in Table~\ref{tab:obs_log}. We assumed a baryon fraction of 15\% within $r_{\rm 200c}$ \cite[e.g.,][]{Eckert22} to estimate the DM mass. Then we assume an NFW mass model for the DM and adopt the DM concentration parameter $c_{\rm 200c}$ from the $c_{\rm 200c}-M_{\rm 200c}$ relation from \cite{Dutton14}, which is consistent with more recent studies \cite[e.g.,][]{Diemer19} and simulations with warm DM halos \citep[e.g.,][]{2016MNRAS.460.1214L}.
The $c_{\rm 200c}$ values range between 4.0--5.3 in our mass range.
While the measured values of $c_{\rm 200c}$ for some clusters are higher (e.g., Virgo and Centaurus, see references in Table~\ref{tab:obs_log}), those fits are for the total mass profiles and can also be affected by limited spatial coverage. 
For Centaurus and Ophiuchus, we used the relevant X-ray temperatures (2.9 keV and 9.1 keV respectively) in the cited works and estimated $M_{\rm 500c}$ from the $M-T$ relation from \cite{Sun09}. For Centaurus, $M_{\rm 200c}$ = 1.39 $M_{\rm 500c}$ for $c_{\rm 200c} = 4.9$. For Ophiuchus, $M_{\rm 200c}$ = 1.44 $M_{\rm 500c}$ for $c_{\rm 200c} = 4.0$.

The individual cluster mass models and $M_{\rm 200c}$ values have significant uncertainty; for example, \cite{2022NatAs...6..936H} quotes $M_{\rm 200c}$ for Coma twice as high as our value in Table~\ref{tab:obs_log}. The random component of this uncertainty is reduced by our use of a sample of clusters.
We discuss the effect of the DM mass uncertainty on the DM decay rate in \S\ref{sect:dis}. 

The model for each cluster is used to generate a projected DM mass map by integrating the mass profile along the line of sight to $1.3 r_{\rm 200c}$ (using the outer radius between $(1-2)r_{\rm 200c}$ results in $<1$\% differences). 
The projected mass map is then smoothed with a Gaussian with $\sigma = 33''$ that corresponds to the XRISM PSF.
We evaluate the projected mass within the Resolve region used for the spectral extraction --- typically the full FOV (again pixel 12 is not in the FOV and is always excluded) with pixel 27 excluded (with more pixels excluded for some observations, see Table~\ref{tab:obs_log}). The resulting projected masses are given in Table~\ref{tab:obs_log}. A $w$ factor is defined for each cluster (see the caption of Table~\ref{tab:obs_log}), which is proportional to the number of expected DM decay photons.
We also calculated the $w$ factor for the Hitomi data of the Perseus cluster \citep{Markevitch17}, with the difference in the rest-frame 3.55 keV effective area accounted for. Our total $w$ factor is 10.2 times the Hitomi value, also with weaker ICM emission on average at 3.4 - 3.8 keV (Table~\ref{tab:obs_log}).
We note that the measured $c_{\rm 200c}$, typically for the total mass profile, is always higher than the value we adopted. If the measured $c_{\rm 200c}$ (available for seven clusters, excluding Coma, Ophiuchus and A2199) are used, the combined $w$ factor for these seven clusters will be 32\% higher, which would decrease our limit on the DM decay rate by 24\%. The uncertainty of the mass modeling is further discussed in Section~\ref{sect:dis}.

\subsection{Spectral stacking}
\label{sect:stacking}

The cluster spectra are stacked in the source rest frame, in order to amplify any line signal common to clusters. 
For that, individual cluster spectra, responses and background models need to be de-redshifted to the cluster rest frame. There are multiple values of the cluster redshift --- the one based on the optical galaxy velocity average and the X-ray redshifts, which in turn can be different for the different regions of the cluster. Which value should be used depends on the goal of the line search. For individual regions, an apparent X-ray redshift is derived from the spectral fit of the XRISM Resolve spectra (without applying the barycentric correction to it). Such an observed X-ray redshift can be applied in reverse to align the atomic lines from the ICM.
The optical redshift (the average for the member galaxies with measured
spectroscopic redshifts, shown in Table~\ref{tab:obs_log}) is likely to be the centroid velocity for a DM emission line.
As optical redshifts are given in the Sun frame, the barycentric velocity corrections were applied in reverse to convert them to the XRISM frame.
In this work, we tried to stack the spectra using both redshift values, which resulted in very small differences for the results, as discussed in \S~\ref{sect:dis}.

To de-redshift the spectra, we scaled the observed photon energies by $(1+z)$, taking care to conserve the number of photons and avoid rounding artifacts. We then used an FTOOL mathpha to coadd the spectra.
The stacked spectrum of the 36 XRISM observations has a total exposure time of 3.748 Ms. We also divide the full sample into the cool and hot subsamples. The cool subsample includes M87, Centaurus, Hydra A and A2199 (all with $M_{\rm 200c} < 10^{14.5}$ M$_{\odot}$), with a total exposure time of 1.259 Ms.
The hot subsample includes Perseus, Coma, A2319, A2029, PKS\,0745--19 and Ophiuchus (all with $M_{\rm 200c} > 10^{14.5}$ M$_{\odot}$), with a total exposure time of 2.489 Ms.

The NXB model consists of a power law and 15 narrow Gaussian lines at 2--12 keV. We also de-redshifted the NXB model for each observation and combined all 36 of them into the final NXB model for the de-redshifted stacked spectrum. The NXB model is used with a diagonal RMF and no ARF, as is the standard approach.
There can also be DM lines from the Milky Way. However, such signals remain undetected from searches of deep X-ray CCD data (see Section~\ref{sect:intro}). As shown in \cite{2012PhRvD..85f3517C,2016PhRvD..93j3512E}, the Milky Way DM line should be more than one order of magnitude fainter than the cluster DM line so it is negligible.

The spectral responses were averaged after adjusting to redshift zero from the redshift assumed for each cluster. The point-source-at-the-aimpoint ARF, accounting for the subarray used for each observation, is calculated in the standard way. Then, the effective areas in the file are replaced with the value in the $z=0$ effective area curve computed using interpolation at $E/(1+z)$, where $E$\/ is the mid-point of each ARF energy bin, and then reduced by the $(1+z)$ stretch factor associated with the de-redshifted energy bin in the stacked spectrum. The de-redshifted RMF for each observation was computed using an RMF parameters CalDB file (Hp Gaussian core component only) as follows. The FWHM for each pixel in each file is replaced with the value calculated at $E/(1+z)$ from interpolation of the FWHM versus energy curve in the standard file. The value of the FWHM is then increased by the $(1+z)$ stretch factor --- as a result, the overall normalization of the line spread function is maintained, but the fraction within an energy bin is reduced.

Because the redshifts in our sample are modest and the resolution is a slowly varying function of the energy, the effect of registration of all clusters to their rest frames results in a reduction in resolution that is less than 0.4 eV (the value for PKS\,0745--19) at $E=3.5$ keV for any cluster. 
The ARF and RMF are matrix-multiplied for each OBSID to generated RSP files. For a check, we applied this technique to the spectra of A2029 and Coma, and correctly recovered the best-fit temperatures, abundances, bulk velocities, and velocity dispersions. The RSP files are then combined using the \texttt{ftaddrmf} in FTOOLS and the weights in Table~\ref{tab:obs_log} based on the projected DM masses.  We also generated the combined response files using two other different weights, one by X-ray counts and the other by count rates. The changes on the final upper limits are always within 4\%. Finally, we tested an extension of the RMF calculation with the exponential tail component (in addition to the Gaussian core) and found that its inclusion had no impact on the line flux constraints, as expected based on its small contribution.

\begin{table}[!t]
    \begin{center}
    \caption{The baseline spectral models for line search}
    \label{tab:baseline}
   \vspace{-0.4cm}
    \begin{tabular}{lll}
    \hline
        E range & 1st bapec & 2nd bapec \\
        \hline\hline
        2.4-4.1 keV & $kT=1.86\pm0.07$ keV & $kT=5.09^{+0.29}_{-0.20}$ keV \\
        690/692 & $Z=0.75\pm0.09$ $Z_{\odot}$    & $Z=0.78^{+0.10}_{-0.08}$ $Z_{\odot}$ \\ 
                & $\sigma=149^{+17}_{-18}$ km/s  & $\sigma=150^{+27}_{-31}$ km/s \\
        3.7-6.5 keV & $kT=2.56^{+0.04}_{-0.07}$ keV & $kT=11.2^{+0.7}_{-0.5}$ keV \\
        1183/1119   & $Z=0.62\pm0.03$ $Z_{\odot}$   & $Z=1.4\pm0.2$ $Z_{\odot}$ \\ 
                    & $\sigma=150\pm13$ km/s  & $\sigma=155^{+73}_{-79}$ km/s \\
         6.3-15.1 keV  & $kT=3.54^{+0.33}_{-0.14}$ keV & $kT=8.89^{+0.33}_{-0.34}$ keV \\
        491/409   & $Z=0.58\pm0.07$ $Z_{\odot}$   & $Z=0.39\pm0.04$ $Z_{\odot}$ \\ 
                    & $\sigma=123\pm21$ km/s  & $\sigma=189\pm42$ km/s \\          
        \hline
    \end{tabular}
    \end{center}
   \vspace{-0.2cm}
Note: the number below the energy range shows the fit C-statistics and the degree of freedom.
\end{table}

\subsection{Searching for unidentified lines}
\label{sect:spec1}

The rest-frame 3--4 keV stacked spectra for the full sample and cool and hot subsamples are shown in Fig.~\ref{fig:spec}, while the full-sample spectrum for the rest of the 2--10 keV band is shown in the Appendix (Fig.~\ref{fig:spe1} and Fig.~\ref{fig:spe2}). We first performed a visual search for emission lines from the stacked spectra in the 2--15 keV range, with the line information from AtomDB\footnote{\url{http://www.atomdb.org/index.php}}. All the emission lines identified have an atomic origin.

We then ran systematic searches for emission/absorption lines in the stacked spectra to look for unidentified lines. Our method is similar to the one adopted in \cite{Markevitch17} and \cite{Tamura19}. We first fit the stacked spectra with a thermal plasma model as the baseline model. Then a Gaussian line model with a fixed width (see the later discussion on the expected line width) is added to represent an additional emission or absorption line at each energy, and limits on its flux are derived as a function of the line energy.
The stacked spectra were fitted using Xspec (v12.15.0d in HEASoft 6.35.2) employing C-statistics \citep{1979ApJ...228..939C}. For the ICM plasma model, we used a velocity-broadened, collisional-equilibrium model (\texttt{bapec}), with the atomic data from AtomDB v3.1.2 \citep{2012ApJ...756..128F}.
To cleanly isolate lines, our method requires a good fit to the continuum.
As a satisfactory fit to the full 2--15 keV range cannot be achieved for the stacked spectra, even with two \texttt{bapec} models, we performed the spectral fits in three separate energy ranges, 2.4--4.1 keV, 3.7--6.5 keV and 6.3--15.1 keV.
With Resolve's high spectral resolution, our line search is the local search.
Those baseline models are listed in Table~\ref{tab:baseline}.
To achieve a good continuum fit in the 6.3--15.1 keV range, we masked the bright ICM Fe He$\alpha$ lines at 6.58--6.72 keV and Ly$\alpha$ lines at 6.89--7.00 keV, because the residuals are seen mainly around those bright lines.

We also attempted to include a multiplicative photoelectric absorption model. However, as the best-fit column density is always consistent with zero, such a component was omitted from the spectral model.
Particularly for the line search around 3.5 keV, we also attempted a simple power law model to fit the rest-frame 3.42--3.83 keV spectra, adding two narrow Gaussian components for Ar and K atomic lines at their database energies. The resulting $3\sigma$ constraints on the additional line are nearly the same as those derived from the above model.

The ICM lines are narrow, with $\sigma\sim 100 - 200$ km/s, determined by the ICM trubulent and random motions \cite[e.g.,][]{XRISMCenI,XRISM2029a,XRISMcomaa,Fujita25}, while a DM decay line should have a higher width that corresponds to the DM particle velocity dispersion. 
We use the cluster mass - velocity dispersion relation from \cite{2013MNRAS.430.2638M} and assume this relation can be used for warm DM particles, such as a keV-mass sterile neutrino.
Then we calculated a weighted $\sigma_{\rm 1D}^2$, with the weights given in Table~\ref{tab:obs_log}. The derived vaues of $\sigma_{\rm 1D}$ for DM from the full sample, the cool subsample and the hot subsample are 950, 620 and 1100 km s$^{-1}$ respectively, which are then assumed in the spectral analysis.

With the above models, we searched for unidentified lines in the 2.5--15 keV range, using a 2.5 eV step for the line energy. At $E< 2.5$ keV, the XRISM/Resolve effective area rapidly drops ($<21$ cm$^{2}$ in the weighted ARF), so the limits are weak. During the search, the \texttt{bapec} temperature, abundance, velocity dispersion, normalization and the Gaussian line normalization were allowed to change. This search was performed for the line widths that correspond to the DM origin ($\sigma_{\rm 1D}=950$ km\,s$^{-1}$) or turbulent ICM origin ($\sigma_{\rm 1D}=160$ km\,s$^{-1}$).
The resulting limits are shown in Fig.~\ref{fig:limits}. The $3\sigma$ upper limit in the $E=3.52-3.60$ keV interval is $\sim 10^{-6}$ photons cm$^{-2}$ s$^{-1}$ for the broad line. 
Because this is not a detection at {\rm some}\/ energy in the broad band, but rather an upper limit that applies to {\rm each}\/ energy bin within the band, the ``look-elsewhere effect'' does not apply to this result, as discussed in \cite{Markevitch17}.
Above 10 keV, our results are limited by the limited statistics and the NXB modeling (e.g., the remaining NXB emission at $\sim$ 10 keV, $\sim$ 12.2 keV and $\sim$ 13.7 keV, which should be improved with more data and the better NXB model in the future.

\section{Discussion and Conclusions}
\label{sect:dis}

We can convert our updated line flux constraints to an upper limit on the DM particle decay rate, knowing the DM mass that we are looking at (Table~\ref{tab:obs_log}). Details can be found in B14. 
Basically, the DM decay rate $\Gamma_{\gamma} = 4 \pi \frac{F_{\rm DM}}{w_{\rm total}} m_{\rm s} Exp_{\rm total}$, where $F_{\rm DM}$ is the observed X-ray limit on the DM line, $w_{\rm total}$ is the total $w$ factor combined, $m_{\rm s}$ is the assumed DM mass and $Exp_{\rm total}$ is total exposure time.
The average $3\sigma$ upper limit on the photon rate in the stacked spectrum, in the rest-frame 3.52--3.60 keV band is 0.95 $\times10^{-6}$ photons cm$^{-2}$ s$^{-1}$, which corresponds to a DM decay rate of 0.97 $\times10^{-27}$ s$^{-1}$, assuming a DM particle mass $m_s=7.1$ keV. This limit is 3--4 times lower than that derived from the Hitomi data for the Perseus cluster \citep{Markevitch17}, but still a factor 5 higher than the corresponding DM decay rate of $\sim 2 \times 10^{-28}$ s$^{-1}$ from the line detected in the XMM-Newton cluster stacking data (B14).
For convenience, the $3\sigma$ upper limits shown in Fig.~\ref{fig:limits} can be approximated by a 4th-order polynomial $f = 10^{-6} f_{1}$ photons s$^{-1}$ cm$^{-2}$ at the 2.5--15 keV, where $f_{1} =$ 7.8774 - 3.5017$E$ + 0.59087 $E^{2}$ - 0.042245 $E^{3}$ + 0.0010943 $E^{4}$
and $E$\/ is the line energy in keV (also shown in Fig.~\ref{fig:limits}).
We can further convert it to the upper limit to the sterile neutrino mixing angle,
$sin^{2} 2\theta / 10^{-10}< 4.1 (E / 3.55)^{-4} f_{1}$.
Our limits are not stronger than some of existing limits based on the CCD data (see e.g., Fig. 4 of \citealt{sicilian22}). However, the XRISM data can resolve the claimed DM line, while the CCD spectra cannot and are subject to different systematic uncertainty from our data.

Increasing or decreasing $M_{\rm 200c}$ of all the clusters by a factor of 1.5 from the values shown in Table~\ref{tab:obs_log} varies the $w$ factor by +18\% or -16\% according to our simulations.
The smaller change downward is related to the slight anti-correlation between concentration parameter and mass. The relatively small change on the $w$ factor is not surprising. The DM concentration parameter in our mass range is nearly constant. With a constant concentration for the NFW profile, increasing $M_{\rm 200c}$ by a factor of 1.5 means increasing the scale radius of the NFW profile by 1.5$^{1/3}$ = 1.14 while keeping the central density the same. For Resolve's small FOV, all observations in this sample, except for the outmost pointing of A2029 (000152000 and 300053010), are well within the scale radius of the DM core, where the density profile is less steep than that beyond the scale radius.
The change in the $w$ factor stems from the increased size and subsequently more projected mass for a more massive cluster.
If the $w$ factor for a single cluster has a random $\sim$ 18\% uncertainty (a reasonable assumption, as the cluster mass models were derived with different data and methods), the total $w$ factor for ten clusters would be less uncertain ($\sim$ 5.7\% if each cluster has the same factor).
A systematic bias may apply to all the mass estimates, e.g., the hydrostatic equilibrium (HSE) mass bias. But the HSE mass bias should be smaller than 50\% \cite[e.g.,][]{2019A&A...621A..40E} and it will make our constraint on the DM decay rate more restrictive.
A sample study has the clear advantage of the smaller effects of the mass uncertainties and the instrument response inaccuracy, compared to results based on single systems.

Eight of ten clusters are cool cores, and the current Resolve observations focus on cool cores, because of their high X-ray brightness. On the other hand, cool core clusters are typically more relaxed and are likely to have higher halo concentrations \citep[e.g.,][]{2023MNRAS.521..790D}. If we keep $M_{\rm 200c}$ the same, higher $c_{\rm 200c}$ would result in larger $w$\/ for Resolve, which would make our limits on the DM decay rate more restrictive. We tested this on Virgo and Ophiuchus, two clusters with the most extreme $c_{\rm 200c}$ values in this sample. For Virgo, if we increase its $c_{\rm 200c}$ from 5.3 to 8.6 that is measured from the total mass profile \citep{Simionescu17} and keep its $M_{\rm 200c}$ the same, its total $w$ increases by 56\%. For Ophiuchus, if we increase its $c_{\rm 200c}$ from 4.0 to 6.0 (see the relation for clusters with similar mass in \citealt{2023MNRAS.521..790D}) and keep its $M_{\rm 200c}$ the same, its total $w$ increases by 39\%. Thus, our limit can be stronger with this effect considered. On the other hand, mis-centering of DM halos within 0.02 $r_{\rm 200c}$ would result in a $\sim$ 10\% overestimate of $w$. Considering these uncertainties, the $w$ factors shown in Table~\ref{tab:obs_log} still result in a conservative estimate of the upper limit on the DM decay rate.

We also used the optical redshifts to perform the correction for both spectra and responses to repeat the stacking and spectral analysis. Because the optical $z$\/ is often different from the X-ray $z$, we have to free the velocity of the \texttt{bapec} component to get the satisfactory fits. For example, the model with two \texttt{bapec} components has the best fit with velocities of --64 km s$^{-1}$ and --81 km s$^{-1}$ for the full sample. In contrast, as expected, the same model applied to the stacked spectrum using the X-ray redshifts has velocities consistent with zero. Thus, we allow the velocity of \texttt{bapec} component to change in the baseline models, when the stacked spectra with the optical redshift are studied. On the other hand, the rest-frame 3.4--3.8 keV range has only weak lines, so the velocities of the \texttt{bapec} components have little impact on the line limits. We repeat the same analysis as above and find the limits in the 3.4--3.8 keV rest-frame band essentially the same.

We also derived the 3.4--3.8 keV band limits for the cool and hot subsamples.
The constraints from the cool subsample are only $\sim$ 10\% worse than those from the full sample. This is due to the combination of two factors. First, the continuum around 3.5 keV in the cool subsample is only $\sim$ 40\% of that in the hot subsample (Fig.\ 1). Second, the expected DM line width for the cool subsample is 56\% of that for the hot subsample. The constraints from the hot subsample are $\sim$ 80\% worse than those from the full sample.

In fact, the above results for different samples are consistent with the following simple estimate for the signal-to-noise ratio (S/N) of the expected DM line signal. Roughly, the DM line S/N $\sim \frac{w A}{\sqrt{f \Delta{}E A t}} = \frac{w / t}{\sqrt{f \Delta{}E}} A^{1/2} t^{1/2}$, where $w$ is the factor as defined in the notes for Table~\ref{tab:obs_log} (also see B14),
or $w = (\sum_{i}{t_i * M_i}$) * (1+z) / $D_{\rm L}^2$, $A$ is the effective area around the observed energy of the DM line, $f$ is the flux density of the cluster continuum around the DM line, $\Delta{}E$ is the width adopted for the line search and $t$ is the total exposure time ($t_i$ is an individual exposure time).
We can take $\Delta{}E$ as the width of the DM line but it should not be smaller than Resolve's energy resolution at $z$.
Here we also assume that the noise is dominated by the cluster ICM emission. We can compare the expected S/N for the cool subsample and the full sample. Assuming the same $A$ (or ignoring the small $z$ difference), the expected S/N for the cool subsample is $\sim$ 6\% higher than that for the full sample, as the smaller exposure is largely compensated for by the smaller $\Delta{}E$ and $f$ ($\sim$ 1.9 times smaller). This is close to what we observe, since the above simple model underestimates the noise. A similar comparison between the cool subsample and the hot subsample suggests that the cool subsample can provide a constraint that is $\sim$ 70\% better, which is again similar to what we observe.

It is useful to study the prospect of future XRISM observations to reach the sensitivity to test the claimed XMM-Newton detection by B14. We want to increase the S/N by a factor of five. For the full sample, that would require an exposure 25 times longer (or 94 Ms). However, the full sample is not optimized for such a search, as it includes clusters that contribute little to the final limit (e.g., PKS\,0745--19). In fact, if one would focus on the cool subsample, the required exposure time would be decreased by $\sim 3$ times (to 1.259 Ms $\times$ 25 = 31.5 Ms).
The cool subsample can be further optimized for the DM line detection, by including
nearby poor clusters and galaxy groups with even weaker X-ray emission around 3.5 keV.

We further study the relationship between the DM mass enclosed within the Resolve FOV (or $M_{\rm XRISM}$, with pixel 27 excluded) and the mass and distance of a cluster. We assume the form $M_{\rm XRISM} \propto M_{\rm 200c}^{a} D_{\rm A}^{b}$, where $D_{\rm A}$ is the angular diameter distance. Based on the results in Table~\ref{tab:obs_log} and more simulations at lower masses, we find that $M_{\rm XRISM} \propto M_{\rm 200c}^{0.4} D_{\rm A}^{1.55}$.
Note that the dependence on $M_{\rm 200c}^{0.4}$ at a fixed distance is consistent with the estimate from the earlier discussion of the uncertainty on the $w$ factor. $b < 2$ is also expected for the DM density gradient.
Thus, $w / t \propto M_{\rm 200c}^{0.4} D_{\rm L}^{-0.45} (1+z)^{-2.1}$.
One can also relate the continuum flux at $E=3.5$ keV to $M_{\rm 200c}$. Such a relation for cluster cores has a large scatter, e.g., due to cool cores vs. non cool cores and central AGN contamination. While detailed simulations should be done, we simply use our sample (the rest-frame 3.4 -- 3.8 keV count rate in Table~\ref{tab:obs_log}) to find an empirical relation between $M_{\rm 200c}$ and $f$ (the continuum flux around 3.5 keV): $f \propto M_{\rm 200c}^{\sim 0.3}$, which indeed comes with a large scatter. The width of the DM line $\Delta{}E \propto M_{\rm 200c}^{0.334}$ comes from \cite{2013MNRAS.430.2638M}.
Thus, we expect the DM line S/N $\propto M_{\rm 200c}^{0.08} D_{\rm L}^{-0.45} (1+z)^{-2.1} A^{1/2} t^{1/2}$.
One can see that the S/N only depends weakly on $M_{\rm 200c}$, so nearby groups and poor clusters are indeed good targets.
The dependence on distance is also not strong so clusters with $z < 0.1$ can all contribute.
Non cool core clusters are also good targets.

Signal from the Milky Way, present in every observation, can also be searched for \citep[e.g.,][]{2016PASJ...68S..31S,dessert20,2024PASJ...76..512F,2025arXiv250304726Y}, which can present stronger constraints on the DM line than clusters, groups and galaxies \citep[e.g.,][]{dessert20}.
While a detailed study to optimize the XRISM observing strategy for detection of the DM line is beyond the scope of this paper, the above estimates suggest that it should be possible to reach the sensitivity similar to the XMM-Newton detection within the XRISM mission lifetime.

\begin{acknowledgments}
We thank the referee for the prompt review and helpful comments.
We gratefully acknowledge the hard work over many years of all of the engineers and scientists who made the XRISM mission possible. Part of this work was performed under the auspices of the U.S. Department of Energy by Lawrence Livermore National Laboratory under Contract DE-AC52-07NA27344. The material is based upon work supported by NASA under award numbers 80GSFC21M0002 and 80GSFC24M0006. This work was supported by JSPS KAKENHI grant numbers JP22H00158, JP22H01268, JP22K03624, JP23H04899, JP21K13963, JP24K00638, JP24K17105, JP21K13958, JP21H01095, JP23K20850, JP24H00253, JP21K03615, JP24K00677, JP20K14491, JP23H00151, JP19K21884, JP20H01947, JP20KK0071, JP23K20239, JP24K00672, JP24K17104, JP24K17093, JP20K04009, JP21H04493, JP20H01946, JP23K13154, JP19K14762, JP20H05857, JP23K03459, and JP25K23398. This work was supported by NASA grant numbers 80NSSC20K0733, 80NSSC18K0978, 80NSSC20K0883, 80NSSC20K0737, 80NSSC24K0678, 80NSSC18K1684, 80NSSC23K0650, and 80NNSC22K1922. LC acknowledges support from NSF award 2205918. CD acknowledges support from STFC through grant ST/T000244/1. LG acknowledges financial support from Canadian Space Agency grant 18XARMSTMA. MS acknowledges the support by the RIKEN Pioneering Project Evolution of Matter in the Universe (r-EMU) and Rikkyo University Special Fund for Research (Rikkyo SFR). AT and the present research are in part supported by the Kagoshima University postdoctoral research program (KU-DREAM). S.U. acknowledges support by Program for Forming Japan's Peak Research Universities (J-PEAKS). SY acknowledges support by the RIKEN SPDR Program. IZ acknowledges partial support from the Alfred P. Sloan Foundation through the Sloan Research Fellowship. 
SE acknowledges the financial contribution from the {\it Bando INAF per la Ricerca Fondamentale 2024} with a {\it Theory Grant} on 
``Constraining the non-thermal pressure in galaxy clusters with high-resolution X-ray spectroscopy'' (1.05.24.05.10).
HRR acknowledges support from an Anne McLaren Fellowship from the University of Nottingham.
NW and CZ are supported by the GACR grant 21-13491X. This work was supported by the JSPS Core-to-Core Program, JPJSCCA20220002. The material is based on work supported by the Strategic Research Center of Saitama University.
\end{acknowledgments}

\vspace{5mm}
\facilities{XRISM(Resolve)}

\software{Xspec \citep{1996ASPC..101...17A},
          HEASoft
}

\appendix
\section{The stacked spectra}
\label{app:spe}

\begin{figure*}
\includegraphics[width=17.0cm]{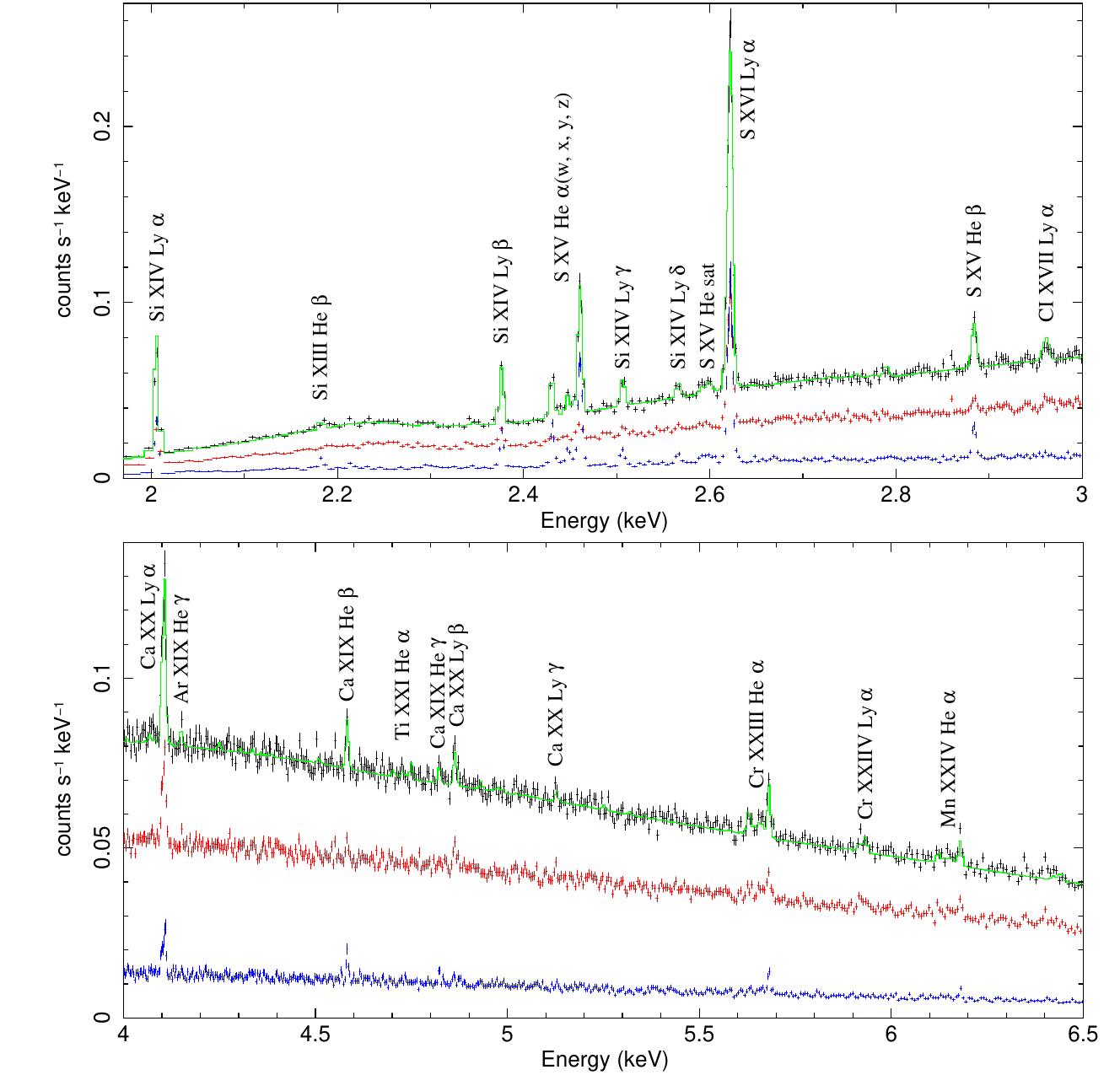}
\vspace{-0.2cm}
\caption{
The stacked spectra in the 1.97--3 keV and 4--6.5 keV ranges, with the best-fit two bapec model shown in the green curve.
The same as Fig.~\ref{fig:spec}, black, red and blue show the full cluster sample, hot subsample and cool subsample respectively.
For clarity, the hot cluster spectrum is lowered by a factor of two and the cool cluster spectrum is lowered by a factor of three.
Detected atomic lines are also labeled.
}
\label{fig:spe1}
\end{figure*}

\begin{figure*}
\includegraphics[width=17.0cm]{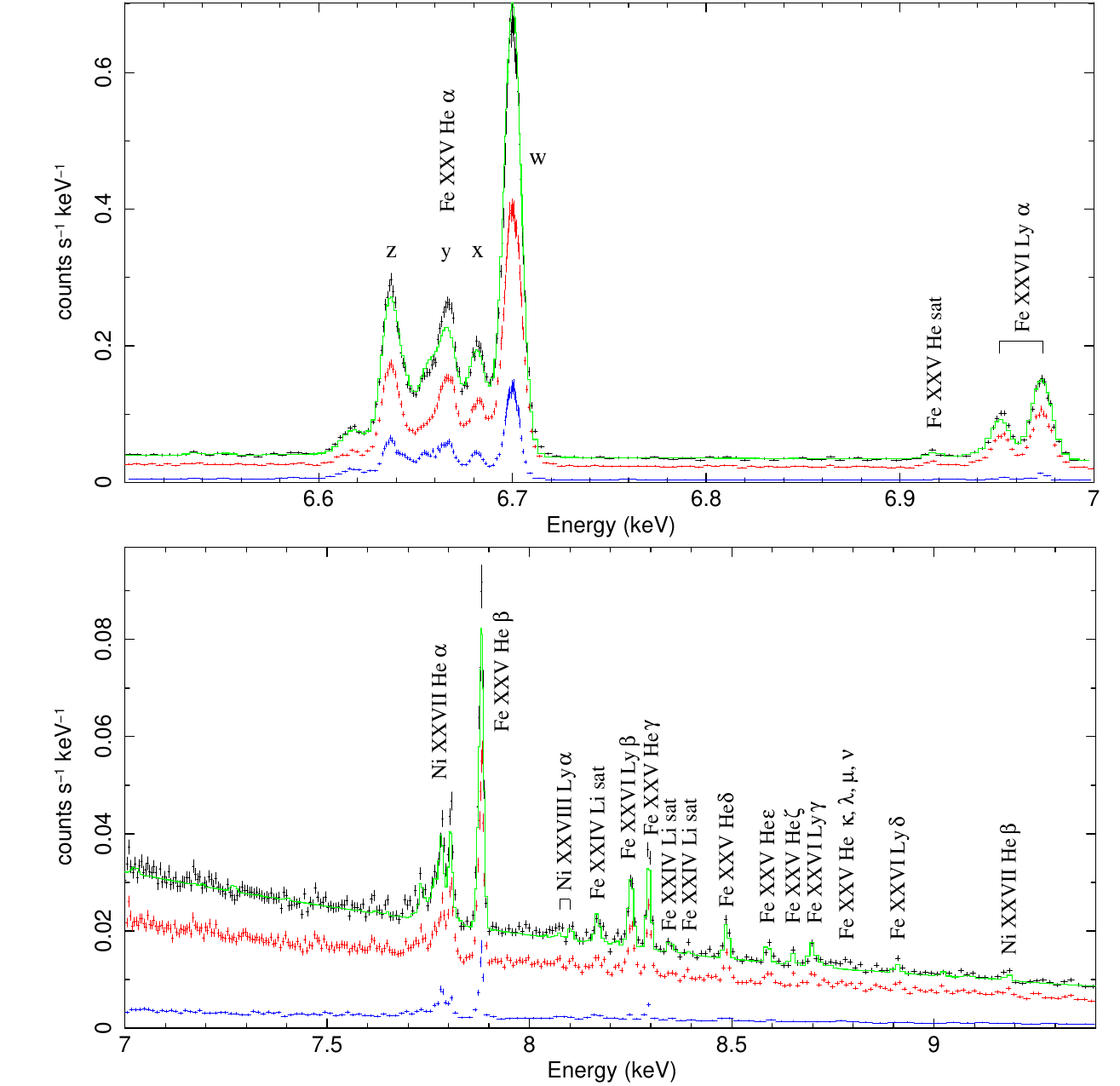}
\vspace{-0.2cm}
\caption{
Similar to Fig.~\ref{fig:spe1} but in the 6.5--7.0 keV and 7.0--9.4 keV ranges.
}
\label{fig:spe2}
\end{figure*}

The stacked spectra from the full sample and two subsamples are shown in Fig.~\ref{fig:spe1} and Fig.~\ref{fig:spe2}. While the goal of the current study is to find any unidentified lines that could be ascribed to DM decay, the spectrum reveals interesting details at the energies of known atomic transitions. A significant excess above the thermal model is seen at $E=8.752-8.773$ keV (at $\sim 2.5\sigma$ significance, see Fig.\ 2). These are the high-$n$\/ transitions of Fe\,XXV, a possible charge exchange signature, earlier detected in Perseus by Hitomi \citep{Hitomi18_atomic}.
There is a $\sim 3\sigma$ detection of the Fe fluorescent line at $E=6.40$ keV, also seen by Hitomi in Perseus \citep{Hitomi18_AGN}. It may come from the X-ray AGN in Perseus, M87 and other systems.
The Fe~XXVI Ly$\alpha_1/\alpha_2$ components at $E=6.95-6.97$ keV show a ratio that differs significantly from 2:1, as reported earlier the XRISM data on Coma and A2029 \citep{XRISMcomaa,XRISM2029a}. These features will be studied in future papers.

\bibliography{references}

\begin{thebibliography}{}
\expandafter\ifx\csname natexlab\endcsname\relax\def\natexlab#1{#1}\fi
\providecommand{\url}[1]{\href{#1}{#1}}
\providecommand{\dodoi}[1]{doi:~\href{http://doi.org/#1}{\nolinkurl{#1}}}
\providecommand{\doeprint}[1]{\href{http://ascl.net/#1}{\nolinkurl{http://ascl.net/#1}}}
\providecommand{\doarXiv}[1]{\href{https://arxiv.org/abs/#1}{\nolinkurl{https://arxiv.org/abs/#1}}}

\bibitem[{{Abazajian} {et~al.}(2001){Abazajian}, {Fuller}, \& {Patel}}]{abazajian01}
{Abazajian}, K., {Fuller}, G.~M., \& {Patel}, M. 2001, \prd, 64, 023501, \dodoi{10.1103/PhysRevD.64.023501}

\bibitem[{{Abazajian}(2017)}]{abazajian17}
{Abazajian}, K.~N. 2017, \physrep, 711, 1, \dodoi{10.1016/j.physrep.2017.10.003}

\bibitem[{{Anderson} {et~al.}(2015){Anderson}, {Churazov}, \& {Bregman}}]{anderson15}
{Anderson}, M.~E., {Churazov}, E., \& {Bregman}, J.~N. 2015, \mnras, 452, 3905, \dodoi{10.1093/mnras/stv1559}

\bibitem[{{Arnaud}(1996)}]{1996ASPC..101...17A}
{Arnaud}, K.~A. 1996, in Astronomical Society of the Pacific Conference Series, Vol. 101, Astronomical Data Analysis Software and Systems V, ed. G.~H. {Jacoby} \& J.~{Barnes}, 17

\bibitem[{{Asplund} {et~al.}(2009){Asplund}, {Grevesse}, {Sauval}, \& {Scott}}]{aspl09}
{Asplund}, M., {Grevesse}, N., {Sauval}, A.~J., \& {Scott}, P. 2009, \araa, 47, 481, \dodoi{10.1146/annurev.astro.46.060407.145222}

\bibitem[{{Bhargava} {et~al.}(2024){Bhargava}, {Giles}, {Romer}, {Jeltema}, {Hollowood}, \& {Hilton}}]{Bhargava24_Zw3146}
{Bhargava}, S., {Giles}, P., {Romer}, K., {et~al.} 2024, Research Notes of the American Astronomical Society, 8, 118, \dodoi{10.3847/2515-5172/ad43e4}

\bibitem[{{Boyarsky} {et~al.}(2019){Boyarsky}, {Drewes}, {Lasserre}, {Mertens}, \& {Ruchayskiy}}]{boyarsky19}
{Boyarsky}, A., {Drewes}, M., {Lasserre}, T., {Mertens}, S., \& {Ruchayskiy}, O. 2019, Progress in Particle and Nuclear Physics, 104, 1, \dodoi{10.1016/j.ppnp.2018.07.004}

\bibitem[{{Boyarsky} {et~al.}(2015){Boyarsky}, {Franse}, {Iakubovskyi}, \& {Ruchayskiy}}]{boyarsky15}
{Boyarsky}, A., {Franse}, J., {Iakubovskyi}, D., \& {Ruchayskiy}, O. 2015, \prl, 115, 161301, \dodoi{10.1103/PhysRevLett.115.161301}

\bibitem[{{Boyarsky} {et~al.}(2014){Boyarsky}, {Ruchayskiy}, {Iakubovskyi}, \& {Franse}}]{boyarsky14}
{Boyarsky}, A., {Ruchayskiy}, O., {Iakubovskyi}, D., \& {Franse}, J. 2014, \prl, 113, 251301, \dodoi{10.1103/PhysRevLett.113.251301}

\bibitem[{{Bulbul} {et~al.}(2014){Bulbul}, {Markevitch}, {Foster}, {Smith}, {Loewenstein}, \& {Randall}}]{Bulbul14}
{Bulbul}, E., {Markevitch}, M., {Foster}, A., {et~al.} 2014, \apj, 789, 13, \dodoi{10.1088/0004-637X/789/1/13}

\bibitem[{{Cappelluti} {et~al.}(2018){Cappelluti}, {Bulbul}, {Foster}, {Natarajan}, {Urry}, {Bautz}, {Civano}, {Miller}, \& {Smith}}]{cappelluti18}
{Cappelluti}, N., {Bulbul}, E., {Foster}, A., {et~al.} 2018, \apj, 854, 179, \dodoi{10.3847/1538-4357/aaaa68}

\bibitem[{{Carlson} {et~al.}(2015){Carlson}, {Jeltema}, \& {Profumo}}]{carlson15}
{Carlson}, E., {Jeltema}, T., \& {Profumo}, S. 2015, \jcap, 2015, 009, \dodoi{10.1088/1475-7516/2015/02/009}

\bibitem[{{Cash}(1979)}]{1979ApJ...228..939C}
{Cash}, W. 1979, \apj, 228, 939, \dodoi{10.1086/156922}

\bibitem[{{Combet} {et~al.}(2012){Combet}, {Maurin}, {Nezri}, {Pointecouteau}, {Hinton}, \& {White}}]{2012PhRvD..85f3517C}
{Combet}, C., {Maurin}, D., {Nezri}, E., {et~al.} 2012, \prd, 85, 063517, \dodoi{10.1103/PhysRevD.85.063517}

\bibitem[{{Darragh-Ford} {et~al.}(2023){Darragh-Ford}, {Mantz}, {Rasia}, {Allen}, {Morris}, {Foster}, {Schmidt}, \& {Wenrich}}]{2023MNRAS.521..790D}
{Darragh-Ford}, E., {Mantz}, A.~B., {Rasia}, E., {et~al.} 2023, \mnras, 521, 790, \dodoi{10.1093/mnras/stad585}

\bibitem[{{Dasgupta} \& {Kopp}(2021)}]{2021PhR...928....1D}
{Dasgupta}, B., \& {Kopp}, J. 2021, \physrep, 928, 1, \dodoi{10.1016/j.physrep.2021.06.002}

\bibitem[{{Dessert} {et~al.}(2024{\natexlab{a}}){Dessert}, {Foster}, {Park}, \& {Safdi}}]{Dessert24}
{Dessert}, C., {Foster}, J.~W., {Park}, Y., \& {Safdi}, B.~R. 2024{\natexlab{a}}, \apj, 964, 185, \dodoi{10.3847/1538-4357/ad2612}

\bibitem[{{Dessert} {et~al.}(2024{\natexlab{b}}){Dessert}, {Ning}, {Rodd}, \& {Safdi}}]{2024PhRvL.132u1002D}
{Dessert}, C., {Ning}, O., {Rodd}, N.~L., \& {Safdi}, B.~R. 2024{\natexlab{b}}, \prl, 132, 211002, \dodoi{10.1103/PhysRevLett.132.211002}

\bibitem[{{Dessert} {et~al.}(2020){Dessert}, {Rodd}, \& {Safdi}}]{dessert20}
{Dessert}, C., {Rodd}, N.~L., \& {Safdi}, B.~R. 2020, Science, 367, 1465, \dodoi{10.1126/science.aaw3772}

\bibitem[{{Diemer} \& {Joyce}(2019)}]{Diemer19}
{Diemer}, B., \& {Joyce}, M. 2019, \apj, 871, 168, \dodoi{10.3847/1538-4357/aafad6}

\bibitem[{{Dodelson} \& {Widrow}(1994)}]{dodelson94}
{Dodelson}, S., \& {Widrow}, L.~M. 1994, \prl, 72, 17, \dodoi{10.1103/PhysRevLett.72.17}

\bibitem[{{Dutton} \& {Macci{\`o}}(2014)}]{Dutton14}
{Dutton}, A.~A., \& {Macci{\`o}}, A.~V. 2014, \mnras, 441, 3359, \dodoi{10.1093/mnras/stu742}

\bibitem[{Eckart {et~al.}(2024)Eckart, Brown, Chiao, Cumbee, Fujimoto, Hell, Hoshino, Ishisaki, Kelley, Kenyon, Kilbourne, Kitamoto, Leutenegger, Lockard, Loewenstein, Magee, Mizumoto, Porter, Sato, Sawada, Shah, Shipman, Sneiderman, Takei, Tsujimoto, de~Vries, Watanabe, Witthoeft, Wolfs, Yamada, \& Yaqoob}]{Eckart2024}
Eckart, M.~E., Brown, G.~V., Chiao, M.~P., {et~al.} 2024, in Space Telescopes and Instrumentation 2024: Ultraviolet to Gamma Ray, ed. J.-W.~A. den Herder, S.~Nikzad, \& K.~Nakazawa, Vol. 13093, International Society for Optics and Photonics (SPIE), 130931P, \dodoi{10.1117/12.3019276}

\bibitem[{{Eckert} {et~al.}(2022){Eckert}, {Ettori}, {Pointecouteau}, {van der Burg}, \& {Loubser}}]{Eckert22}
{Eckert}, D., {Ettori}, S., {Pointecouteau}, E., {van der Burg}, R.~F.~J., \& {Loubser}, S.~I. 2022, \aap, 662, A123, \dodoi{10.1051/0004-6361/202142507}

\bibitem[{{Eckert} {et~al.}(2019){Eckert}, {Ghirardini}, {Ettori}, {Rasia}, {Biffi}, {Pointecouteau}, {Rossetti}, {Molendi}, {Vazza}, {Gastaldello}, {Gaspari}, {De Grandi}, {Ghizzardi}, {Bourdin}, {Tchernin}, \& {Roncarelli}}]{2019A&A...621A..40E}
{Eckert}, D., {Ghirardini}, V., {Ettori}, S., {et~al.} 2019, \aap, 621, A40, \dodoi{10.1051/0004-6361/201833324}

\bibitem[{{Ettori} {et~al.}(2019){Ettori}, {Ghirardini}, {Eckert}, {Pointecouteau}, {Gastaldello}, {Sereno}, {Gaspari}, {Ghizzardi}, {Roncarelli}, \& {Rossetti}}]{Ettori19}
{Ettori}, S., {Ghirardini}, V., {Eckert}, D., {et~al.} 2019, \aap, 621, A39, \dodoi{10.1051/0004-6361/201833323}

\bibitem[{{Evans} {et~al.}(2016){Evans}, {Sanders}, \& {Geringer-Sameth}}]{2016PhRvD..93j3512E}
{Evans}, N.~W., {Sanders}, J.~L., \& {Geringer-Sameth}, A. 2016, \prd, 93, 103512, \dodoi{10.1103/PhysRevD.93.103512}

\bibitem[{{Foster} {et~al.}(2012){Foster}, {Ji}, {Smith}, \& {Brickhouse}}]{2012ApJ...756..128F}
{Foster}, A.~R., {Ji}, L., {Smith}, R.~K., \& {Brickhouse}, N.~S. 2012, \apj, 756, 128, \dodoi{10.1088/0004-637X/756/2/128}

\bibitem[{{Foster} {et~al.}(2021){Foster}, {Kongsore}, {Dessert}, {Park}, {Rodd}, {Cranmer}, \& {Safdi}}]{foster21}
{Foster}, J.~W., {Kongsore}, M., {Dessert}, C., {et~al.} 2021, \prl, 127, 051101, \dodoi{10.1103/PhysRevLett.127.051101}

\bibitem[{{Fujita} {et~al.}(2025){Fujita}, {Fukushima}, {Sato}, {Fukazawa}, \& {Kondo}}]{Fujita25}
{Fujita}, Y., {Fukushima}, K., {Sato}, K., {Fukazawa}, Y., \& {Kondo}, M. 2025, \pasj, \dodoi{10.1093/pasj/psaf089}

\bibitem[{{Fujita} {et~al.}(2008){Fujita}, {Hayashida}, {Nagai}, {Inoue}, {Matsumoto}, {Okabe}, {Reiprich}, {Sarazin}, \& {Takizawa}}]{2008PASJ...60.1133F}
{Fujita}, Y., {Hayashida}, K., {Nagai}, M., {et~al.} 2008, \pasj, 60, 1133, \dodoi{10.1093/pasj/60.5.1133}

\bibitem[{{Fukuichi} {et~al.}(2024){Fukuichi}, {Kitamoto}, {Sawada}, \& {Tamura}}]{2024PASJ...76..512F}
{Fukuichi}, M., {Kitamoto}, S., {Sawada}, M., \& {Tamura}, T. 2024, \pasj, 76, 512, \dodoi{10.1093/pasj/psae027}

\bibitem[{{Gu} {et~al.}(2015){Gu}, {Kaastra}, {Raassen}, {Mullen}, {Cumbee}, {Lyons}, \& {Stancil}}]{gu15}
{Gu}, L., {Kaastra}, J., {Raassen}, A.~J.~J., {et~al.} 2015, \aap, 584, L11, \dodoi{10.1051/0004-6361/201527634}

\bibitem[{{Hitomi Collaboration} {et~al.}(2017){Hitomi Collaboration}, {Aharonian}, {Akamatsu}, {Akimoto}, {Allen}, {Angelini}, {Arnaud}, {Audard}, {Awaki}, {Axelsson}, {Bamba}, {Bautz}, {Blandford}, {Bulbul}, {Brenneman}, {Brown}, {Cackett}, {Chernyakova}, {Chiao}, {Coppi}, {Costantini}, {de Plaa}, {den Herder}, {Done}, {Dotani}, {Ebisawa}, {Eckart}, {Enoto}, {Ezoe}, {Fabian}, {Ferrigno}, {Foster}, {Fujimoto}, {Fukazawa}, {Furuzawa}, {Galeazzi}, {Gallo}, {Gandhi}, {Giustini}, {Goldwurm}, {Gu}, {Guainazzi}, {Haba}, {Hagino}, {Hamaguchi}, {Harrus}, {Hatsukade}, {Hayashi}, {Hayashi}, {Hayashida}, {Hiraga}, {Hornschemeier}, {Hoshino}, {Hughes}, {Ichinohe}, {Iizuka}, {Inoue}, {Inoue}, {Inoue}, {Ishibashi}, {Ishida}, {Ishikawa}, {Ishisaki}, {Itoh}, {Iwai}, {Iyomoto}, {Kaastra}, {Kallman}, {Kamae}, {Kara}, {Kataoka}, {Katsuda}, {Katsuta}, {Kawaharada}, {Kawai}, {Kelley}, {Khangulyan}, {Kilbourne}, {King}, {Kitaguchi}, {Kitamoto}, {Kitayama}, {Kohmura}, {Kokubun}, {Koyama}, {Koyama}, {Kretschmar}, {Krimm}, {Kubota},
  {Kunieda}, {Laurent}, {Lebrun}, {Lee}, {Leutenegger}, {Limousin}, {Loewenstein}, {Long}, {Lumb}, {Madejski}, {Maeda}, {Maier}, {Makishima}, {Markevitch}, {Matsumoto}, {Matsushita}, {McCammon}, {McNamara}, {Mehdipour}, {Miller}, {Miller}, {Mineshige}, {Mitsuda}, {Mitsuishi}, {Miyazawa}, {Mizuno}, {Mori}, {Mori}, {Moseley}, {Mukai}, {Murakami}, {Murakami}, {Mushotzky}, {Nakagawa}, {Nakajima}, {Nakamori}, {Nakano}, {Nakashima}, {Nakazawa}, {Nobukawa}, {Nobukawa}, {Noda}, {Nomachi}, {O' Dell}, {Odaka}, {Ohashi}, {Ohno}, {Okajima}, {Ota}, {Ozaki}, {Paerels}, {Paltani}, {Parmar}, {Petre}, {Pinto}, {Pohl}, {Porter}, {Pottschmidt}, {Ramsey}, {Reynolds}, {Russell}, {Safi-Harb}, {Saito}, {Sakai}, {Sameshima}, {Sasaki}, {Sato}, {Sato}, {Sato}, {Sawada}, {Schartel}, {Serlemitsos}, {Seta}, {Shidatsu}, {Simionescu}, {Smith}, {Soong}, {Stawarz}, {Sugawara}, {Sugita}, {Szymkowiak}, {Tajima}, {Takahashi}, {Takahashi}, {Takeda}, {Takei}, {Tamagawa}, {Tamura}, {Tamura}, {Tanaka}, {Tanaka}, {Tanaka}, {Tashiro}, {Tawara},
  {Terada}, {Terashima}, {Tombesi}, {Tomida}, {Tsuboi}, {Tsujimoto}, {Tsunemi}, {Tsuru}, {Uchida}, {Uchiyama}, {Uchiyama}, {Ueda}, {Ueda}, {Ueno}, {Uno}, {Urry}, {Ursino}, \& {de Vries}}]{Markevitch17}
{Hitomi Collaboration}, {Aharonian}, F.~A., {Akamatsu}, H., {et~al.} 2017, \apjl, 837, L15, \dodoi{10.3847/2041-8213/aa61fa}

\bibitem[{{Hitomi Collaboration} {et~al.}(2018{\natexlab{a}}){Hitomi Collaboration}, {Aharonian}, {Akamatsu}, {Akimoto}, {Allen}, {Angelini}, {Audard}, {Awaki}, {Axelsson}, {Bamba}, {Bautz}, {Blandford}, {Brenneman}, {Brown}, {Bulbul}, {Cackett}, {Chernyakova}, {Chiao}, {Coppi}, {Costantini}, {de Plaa}, {de Vries}, {den Herder}, {Done}, {Dotani}, {Ebisawa}, {Eckart}, {Enoto}, {Ezoe}, {Fabian}, {Ferrigno}, {Foster}, {Fujimoto}, {Fukazawa}, {Furuzawa}, {Galeazzi}, {Gallo}, {Gandhi}, {Giustini}, {Goldwurm}, {Gu}, {Guainazzi}, {Haba}, {Hagino}, {Hamaguchi}, {Harrus}, {Hatsukade}, {Hayashi}, {Hayashi}, {Hayashida}, {Hell}, {Hiraga}, {Hornschemeier}, {Hoshino}, {Hughes}, {Ichinohe}, {Iizuka}, {Inoue}, {Inoue}, {Ishida}, {Ishikawa}, {Ishisaki}, {Iwai}, {Kaastra}, {Kallman}, {Kamae}, {Kataoka}, {Katsuda}, {Kawai}, {Kelley}, {Kilbourne}, {Kitaguchi}, {Kitamoto}, {Kitayama}, {Kohmura}, {Kokubun}, {Koyama}, {Koyama}, {Kretschmar}, {Krimm}, {Kubota}, {Kunieda}, {Laurent}, {Lee}, {Leutenegger}, {Limousin}, {Loewenstein},
  {Long}, {Lumb}, {Madejski}, {Maeda}, {Maier}, {Makishima}, {Markevitch}, {Matsumoto}, {Matsushita}, {McCammon}, {McNamara}, {Mehdipour}, {Miller}, {Miller}, {Mineshige}, {Mitsuda}, {Mitsuishi}, {Miyazawa}, {Mizuno}, {Mori}, {Mori}, {Mukai}, {Murakami}, {Mushotzky}, {Nakagawa}, {Nakajima}, {Nakamori}, {Nakashima}, {Nakazawa}, {Nobukawa}, {Nobukawa}, {Noda}, {Odaka}, {Ohashi}, {Ohno}, {Okajima}, {Ota}, {Ozaki}, {Paerels}, {Paltani}, {Petre}, {Pinto}, {Porter}, {Pottschmidt}, {Reynolds}, {Safi-Harb}, {Saito}, {Sakai}, {Sasaki}, {Sato}, {Sato}, {Sato}, {Sawada}, {Schartel}, {Serlemtsos}, {Seta}, {Shidatsu}, {Simionescu}, {Smith}, {Soong}, {Stawarz}, {Sugawara}, {Sugita}, {Szymkowiak}, {Tajima}, {Takahashi}, {Takahashi}, {Takeda}, {Takei}, {Tamagawa}, {Tamura}, {Tanaka}, {Tanaka}, {Tanaka}, {Tashiro}, {Tawara}, {Terada}, {Terashima}, {Tombesi}, {Tomida}, {Tsuboi}, {Tsujimoto}, {Tsunemi}, {Tsuru}, {Uchida}, {Uchiyama}, {Uchiyama}, {Ueda}, {Ueda}, {Uno}, {Urry}, {Ursino}, {Watanabe}, {Werner}, {Wilkins},
  {Williams}, {Yamada}, {Yamaguchi}, {Yamaoka}, {Yamasaki}, {Yamauchi}, {Yamauchi}, {Yaqoob}, {Yatsu}, {Yonetoku}, {Zhuravleva}, {Zoghbi}, \& {Raassen}}]{Hitomi18_atomic}
{Hitomi Collaboration}, {Aharonian}, F., {Akamatsu}, H., {et~al.} 2018{\natexlab{a}}, \pasj, 70, 12, \dodoi{10.1093/pasj/psx156}

\bibitem[{{Hitomi Collaboration} {et~al.}(2018{\natexlab{b}}){Hitomi Collaboration}, {Aharonian}, {Akamatsu}, {Akimoto}, {Allen}, {Angelini}, {Audard}, {Awaki}, {Axelsson}, {Bamba}, {Bautz}, {Blandford}, {Brenneman}, {Brown}, {Bulbul}, {Cackett}, {Chernyakova}, {Chiao}, {Coppi}, {Costantini}, {de Plaa}, {de Vries}, {den Herder}, {Done}, {Dotani}, {Ebisawa}, {Eckart}, {Enoto}, {Ezoe}, {Fabian}, {Ferrigno}, {Foster}, {Fujimoto}, {Fukazawa}, {Furuzawa}, {Galeazzi}, {Gallo}, {Gandhi}, {Giustini}, {Goldwurm}, {Gu}, {Guainazzi}, {Haba}, {Hagino}, {Hamaguchi}, {Harrus}, {Hatsukade}, {Hayashi}, {Hayashi}, {Hayashida}, {Hiraga}, {Hornschemeier}, {Hoshino}, {Hughes}, {Ichinohe}, {Iizuka}, {Inoue}, {Inoue}, {Ishida}, {Ishikawa}, {Ishisaki}, {Iwai}, {Kaastra}, {Kallman}, {Kamae}, {Kataoka}, {Katsuda}, {Kawai}, {Kelley}, {Kilbourne}, {Kitaguchi}, {Kitamoto}, {Kitayama}, {Kohmura}, {Kokubun}, {Koyama}, {Koyama}, {Kretschmar}, {Krimm}, {Kubota}, {Kunieda}, {Laurent}, {Lee}, {Leutenegger}, {Limousin}, {Loewenstein}, {Long},
  {Lumb}, {Madejski}, {Maeda}, {Maier}, {Makishima}, {Markevitch}, {Matsumoto}, {Matsushita}, {McCammon}, {McNamara}, {Mehdipour}, {Miller}, {Miller}, {Mineshige}, {Mitsuda}, {Mitsuishi}, {Miyazawa}, {Mizuno}, {Mori}, {Mori}, {Mukai}, {Murakami}, {Mushotzky}, {Nakagawa}, {Nakajima}, {Nakamori}, {Nakashima}, {Nakazawa}, {Nobukawa}, {Nobukawa}, {Noda}, {Odaka}, {Ohashi}, {Ohno}, {Okajima}, {Ota}, {Ozaki}, {Paerels}, {Paltani}, {Petre}, {Pinto}, {Porter}, {Pottschmidt}, {Reynolds}, {Safi-Harb}, {Saito}, {Sakai}, {Sasaki}, {Sato}, {Sato}, {Sato}, {Sawada}, {Schartel}, {Serlemitsos}, {Seta}, {Shidatsu}, {Simionescu}, {Smith}, {Soong}, {Stawarz}, {Sugawara}, {Sugita}, {Szymkowiak}, {Tajima}, {Takahashi}, {Takahashi}, {Takeda}, {Takei}, {Tamagawa}, {Tamura}, {Tanaka}, {Tanaka}, {Tanaka}, {Tashiro}, {Tawara}, {Terada}, {Terashima}, {Tombesi}, {Tomida}, {Tsuboi}, {Tsujimoto}, {Tsunemi}, {Tsuru}, {Uchida}, {Uchiyama}, {Uchiyama}, {Ueda}, {Ueda}, {Uno}, {Urry}, {Ursino}, {Watanabe}, {Werner}, {Wilkins}, {Williams},
  {Yamada}, {Yamaguchi}, {Yamaoka}, {Yamasaki}, {Yamauchi}, {Yamauchi}, {Yaqoob}, {Yatsu}, {Yonetoku}, {Zhuravleva}, {Zoghbi}, \& {Kawamuro}}]{Hitomi18_AGN}
---. 2018{\natexlab{b}}, \pasj, 70, 13, \dodoi{10.1093/pasj/psx147}

\bibitem[{{Ho} {et~al.}(2022){Ho}, {Ntampaka}, {Rau}, {Chen}, {Lansberry}, {Ruehle}, \& {Trac}}]{2022NatAs...6..936H}
{Ho}, M., {Ntampaka}, M., {Rau}, M.~M., {et~al.} 2022, Nature Astronomy, 6, 936, \dodoi{10.1038/s41550-022-01711-1}

\bibitem[{{Hofmann} \& {Wegg}(2019)}]{2019A&A...625L...7H}
{Hofmann}, F., \& {Wegg}, C. 2019, \aap, 625, L7, \dodoi{10.1051/0004-6361/201935561}

\bibitem[{{Ishisaki} {et~al.}(2022){Ishisaki}, {Kelley}, {Awaki}, {Balleza}, {Barnstable}, {Bialas}, {Boissay-Malaquin}, {Brown}, {Canavan}, {Cumbee}, {Carnahan}, {Chiao}, {Comber}, {Costantini}, {den Herder}, {Dercksen}, {de Vries}, {DiPirro}, {Eckart}, {Ezoe}, {Ferrigno}, {Fujimoto}, {Gorter}, {Graham}, {Grim}, {Hartz}, {Hayakawa}, {Hayashi}, {Hell}, {Hoshino}, {Ichinohe}, {Ishida}, {Ishikawa}, {James}, {Kenyon}, {Kilbourne}, {Kimball}, {Kitamoto}, {Leutenegger}, {Maeda}, {McCammon}, {Miko}, {Mizumoto}, {Okajima}, {Okamoto}, {Paltani}, {Porter}, {Sato}, {Sato}, {Sawada}, {Shinozaki}, {Shipman}, {Shirron}, {Sneiderman}, {Soong}, {Szymkiewicz}, {Szymkowiak}, {Takei}, {Tamura}, {Tsujimoto}, {Uchida}, {Wasserzug}, {Witthoeft}, {Wolfs}, {Yamada}, \& {Yasuda}}]{Ishisaki2022_Resolve}
{Ishisaki}, Y., {Kelley}, R.~L., {Awaki}, H., {et~al.} 2022, in Society of Photo-Optical Instrumentation Engineers (SPIE) Conference Series, Vol. 12181, Space Telescopes and Instrumentation 2022: Ultraviolet to Gamma Ray, ed. J.-W.~A. {den Herder}, S.~{Nikzad}, \& K.~{Nakazawa}, 121811S, \dodoi{10.1117/12.2630654}

\bibitem[{{Jeltema} \& {Profumo}(2015)}]{jeltema15}
{Jeltema}, T., \& {Profumo}, S. 2015, \mnras, 450, 2143, \dodoi{10.1093/mnras/stv768}

\bibitem[{{Kilbourne} {et~al.}(2018){Kilbourne}, {Adams}, {Brekosky}, {Chervenak}, {Chiao}, {Eckart}, {Figueroa-Feliciano}, {Galeazzi}, {Grein}, {Jhabvala}, {Kelly}, {Leutenegger}, {McCammon}, {Scott Porter}, {Szymkowiak}, {Watanabe}, \& {Zhao}}]{2018JATIS...4a1214K}
{Kilbourne}, C.~A., {Adams}, J.~S., {Brekosky}, R.~P., {et~al.} 2018, Journal of Astronomical Telescopes, Instruments, and Systems, 4, 011214, \dodoi{10.1117/1.JATIS.4.1.011214}

\bibitem[{{Lovell}(2023)}]{2023MNRAS.524.6345L}
{Lovell}, M.~R. 2023, \mnras, 524, 6345, \dodoi{10.1093/mnras/stad2237}

\bibitem[{{Ludlow} {et~al.}(2016){Ludlow}, {Bose}, {Angulo}, {Wang}, {Hellwing}, {Navarro}, {Cole}, \& {Frenk}}]{2016MNRAS.460.1214L}
{Ludlow}, A.~D., {Bose}, S., {Angulo}, R.~E., {et~al.} 2016, \mnras, 460, 1214, \dodoi{10.1093/mnras/stw1046}

\bibitem[{{Malyshev} {et~al.}(2014){Malyshev}, {Neronov}, \& {Eckert}}]{malyshev14}
{Malyshev}, D., {Neronov}, A., \& {Eckert}, D. 2014, \prd, 90, 103506, \dodoi{10.1103/PhysRevD.90.103506}

\bibitem[{{Mirakhor} \& {Walker}(2020)}]{2020MNRAS.497.3943M}
{Mirakhor}, M.~S., \& {Walker}, S.~A. 2020, \mnras, 497, 3943, \dodoi{10.1093/mnras/staa2204}

\bibitem[{{Munari} {et~al.}(2013){Munari}, {Biviano}, {Borgani}, {Murante}, \& {Fabjan}}]{2013MNRAS.430.2638M}
{Munari}, E., {Biviano}, A., {Borgani}, S., {Murante}, G., \& {Fabjan}, D. 2013, \mnras, 430, 2638, \dodoi{10.1093/mnras/stt049}

\bibitem[{{Neronov} {et~al.}(2016){Neronov}, {Malyshev}, \& {Eckert}}]{2016PhRvD..94l3504N}
{Neronov}, A., {Malyshev}, D., \& {Eckert}, D. 2016, \prd, 94, 123504, \dodoi{10.1103/PhysRevD.94.123504}

\bibitem[{{Perez} {et~al.}(2017){Perez}, {Ng}, {Beacom}, {Hersh}, {Horiuchi}, \& {Krivonos}}]{2017PhRvD..95l3002P}
{Perez}, K., {Ng}, K. C.~Y., {Beacom}, J.~F., {et~al.} 2017, \prd, 95, 123002, \dodoi{10.1103/PhysRevD.95.123002}

\bibitem[{{Planck Collaboration} {et~al.}(2013){Planck Collaboration}, {Ade}, {Aghanim}, {Arnaud}, {Ashdown}, {Atrio-Barandela}, {Aumont}, {Baccigalupi}, {Balbi}, {Banday}, {Barreiro}, {Bartlett}, {Battaner}, {Benabed}, {Beno{\^\i}t}, {Bernard}, {Bersanelli}, {Bikmaev}, {B{\"o}hringer}, {Bonaldi}, {Bond}, {Borrill}, {Bouchet}, {Bourdin}, {Brown}, {Brown}, {Burenin}, {Burigana}, {Cabella}, {Cardoso}, {Carvalho}, {Catalano}, {Cay{\'o}n}, {Chiang}, {Chon}, {Christensen}, {Churazov}, {Clements}, {Colafrancesco}, {Colombo}, {Coulais}, {Crill}, {Cuttaia}, {Da Silva}, {Dahle}, {Danese}, {Davis}, {de Bernardis}, {de Gasperis}, {de Rosa}, {de Zotti}, {Delabrouille}, {D{\'e}mocl{\`e}s}, {D{\'e}sert}, {Dickinson}, {Diego}, {Dolag}, {Dole}, {Donzelli}, {Dor{\'e}}, {D{\"o}rl}, {Douspis}, {Dupac}, {En{\ss}lin}, {Eriksen}, {Finelli}, {Flores-Cacho}, {Forni}, {Frailis}, {Franceschi}, {Frommert}, {Galeotta}, {Ganga}, {G{\'e}nova-Santos}, {Giard}, {Gilfanov}, {Gonz{\'a}lez-Nuevo}, {G{\'o}rski}, {Gregorio}, {Gruppuso},
  {Hansen}, {Harrison}, {Henrot-Versill{\'e}}, {Hern{\'a}ndez-Monteagudo}, {Hildebrandt}, {Hivon}, {Hobson}, {Holmes}, {Hornstrup}, {Hovest}, {Huffenberger}, {Hurier}, {Jaffe}, {Jagemann}, {Jones}, {Juvela}, {Keih{\"a}nen}, {Khamitov}, {Kneissl}, {Knoche}, {Knox}, {Kunz}, {Kurki-Suonio}, {Lagache}, {L{\"a}hteenm{\"a}ki}, {Lamarre}, {Lasenby}, {Lawrence}, {Le Jeune}, {Leonardi}, {Lilje}, {Linden-V{\o}rnle}, {L{\'o}pez-Caniego}, {Lubin}, {Mac{\'\i}as-P{\'e}rez}, {Maffei}, {Maino}, {Mandolesi}, {Maris}, {Marleau}, {Mart{\'\i}nez-Gonz{\'a}lez}, {Masi}, {Massardi}, {Matarrese}, {Matthai}, {Mazzotta}, {Mei}, {Melchiorri}, {Melin}, {Mendes}, {Mennella}, {Mitra}, {Miville-Desch{\^e}nes}, {Moneti}, {Montier}, {Morgante}, {Munshi}, {Murphy}, {Naselsky}, {Natoli}, {N{\o}rgaard-Nielsen}, {Noviello}, {Novikov}, {Novikov}, {Osborne}, {Pajot}, {Paoletti}, {Perdereau}, {Perrotta}, {Piacentini}, {Piat}, {Pierpaoli}, {Piffaretti}, {Plaszczynski}, {Pointecouteau}, {Polenta}, {Ponthieu}, {Popa}, {Poutanen}, {Pratt}, {Prunet},
  {Puget}, {Rachen}, {Rebolo}, {Reinecke}, {Remazeilles}, {Renault}, {Ricciardi}, {Riller}, {Ristorcelli}, {Rocha}, {Roman}, {Rosset}, {Rossetti}, {Rubi{\~n}o-Mart{\'\i}n}, {Rudnick}, {Rusholme}, {Sandri}, {Savini}, {Schaefer}, {Scott}, {Smoot}, {Stivoli}, {Sudiwala}, {Sunyaev}, {Sutton}, {Suur-Uski}, {Sygnet}, {Tauber}, {Terenzi}, {Toffolatti}, {Tomasi}, {Tristram}, {Tuovinen}, {T{\"u}rler}, {Umana}, {Valenziano}, {Van Tent}, {Varis}, \& {Vielva}}]{PlanckComa}
{Planck Collaboration}, {Ade}, P.~A.~R., {Aghanim}, N., {et~al.} 2013, \aap, 554, A140, \dodoi{10.1051/0004-6361/201220247}

\bibitem[{Porter {et~al.}(2024)Porter, Kilbourne, Chiao, Cumbee, Eckart, Fujimoto, Ishisaki, Kanemaru, Kelley, Leutenegger, Maeda, Mizumoto, Sato, Sawada, Sneiderman, Takei, Tsujimoto, Uchida, Watanabe, \& Yamada}]{Porter2024}
Porter, F.~S., Kilbourne, C.~A., Chiao, M., {et~al.} 2024, in Space Telescopes and Instrumentation 2024: Ultraviolet to Gamma Ray, ed. J.-W.~A. den Herder, S.~Nikzad, \& K.~Nakazawa, Vol. 13093, International Society for Optics and Photonics (SPIE), 130931K, \dodoi{10.1117/12.3018882}

\bibitem[{{Roach} {et~al.}(2020){Roach}, {Ng}, {Perez}, {Beacom}, {Horiuchi}, {Krivonos}, \& {Wik}}]{roach20}
{Roach}, B.~M., {Ng}, K. C.~Y., {Perez}, K., {et~al.} 2020, \prd, 101, 103011, \dodoi{10.1103/PhysRevD.101.103011}

\bibitem[{{Roach} {et~al.}(2023){Roach}, {Rossland}, {Ng}, {Perez}, {Beacom}, {Grefenstette}, {Horiuchi}, {Krivonos}, \& {Wik}}]{roach23}
{Roach}, B.~M., {Rossland}, S., {Ng}, K. C.~Y., {et~al.} 2023, \prd, 107, 023009, \dodoi{10.1103/PhysRevD.107.023009}

\bibitem[{{Rose} {et~al.}(2025){Rose}, {McNamara}, {Meunier}, {Fabian}, {Russell}, {Nulsen}, {Dizdar}, {Heckman}, {McDonald}, {Markevitch}, {Paerels}, {Simionescu}, {Werner}, {Coil}, {Hodges-Kluck}, {Miller}, \& {Wise}}]{Rose25}
{Rose}, T., {McNamara}, B.~R., {Meunier}, J., {et~al.} 2025, arXiv e-prints, arXiv:2505.01494, \dodoi{10.48550/arXiv.2505.01494}

\bibitem[{{Sarkar} {et~al.}(2025){Sarkar}, {Miller}, {Ota}, {Kilbourne}, {McNamara}, {Sun}, {Lovisari}, {Ettori}, {Eckert}, {Szymkowiak}, {Bartalesi}, \& {Loewenstein}}]{XRISM2029c}
{Sarkar}, A., {Miller}, E., {Ota}, N., {et~al.} 2025, arXiv e-prints, arXiv:2508.04958, \dodoi{10.48550/arXiv.2508.04958}

\bibitem[{{Sekiya} {et~al.}(2016){Sekiya}, {Yamasaki}, \& {Mitsuda}}]{2016PASJ...68S..31S}
{Sekiya}, N., {Yamasaki}, N.~Y., \& {Mitsuda}, K. 2016, \pasj, 68, S31, \dodoi{10.1093/pasj/psv081}

\bibitem[{{Shah} {et~al.}(2016){Shah}, {Dobrodey}, {Bernitt}, {Steinbr{\"u}gge}, {Crespo L{\'o}pez-Urrutia}, {Gu}, \& {Kaastra}}]{shah16}
{Shah}, C., {Dobrodey}, S., {Bernitt}, S., {et~al.} 2016, \apj, 833, 52, \dodoi{10.3847/1538-4357/833/1/52}

\bibitem[{{Sicilian} {et~al.}(2020){Sicilian}, {Cappelluti}, {Bulbul}, {Civano}, {Moscetti}, \& {Reynolds}}]{sicilian20}
{Sicilian}, D., {Cappelluti}, N., {Bulbul}, E., {et~al.} 2020, \apj, 905, 146, \dodoi{10.3847/1538-4357/abbee9}

\bibitem[{{Sicilian} {et~al.}(2022){Sicilian}, {Lopez}, {Moscetti}, {Bulbul}, \& {Cappelluti}}]{sicilian22}
{Sicilian}, D., {Lopez}, D., {Moscetti}, M., {Bulbul}, E., \& {Cappelluti}, N. 2022, \apj, 941, 2, \dodoi{10.3847/1538-4357/ac9fcf}

\bibitem[{{Simionescu} {et~al.}(2017){Simionescu}, {Werner}, {Mantz}, {Allen}, \& {Urban}}]{Simionescu17}
{Simionescu}, A., {Werner}, N., {Mantz}, A., {Allen}, S.~W., \& {Urban}, O. 2017, \mnras, 469, 1476, \dodoi{10.1093/mnras/stx919}

\bibitem[{{Simionescu} {et~al.}(2011){Simionescu}, {Allen}, {Mantz}, {Werner}, {Takei}, {Morris}, {Fabian}, {Sanders}, {Nulsen}, {George}, \& {Taylor}}]{2011Sci...331.1576S}
{Simionescu}, A., {Allen}, S.~W., {Mantz}, A., {et~al.} 2011, Science, 331, 1576, \dodoi{10.1126/science.1200331}

\bibitem[{{Sun} {et~al.}(2009){Sun}, {Voit}, {Donahue}, {Jones}, {Forman}, \& {Vikhlinin}}]{Sun09}
{Sun}, M., {Voit}, G.~M., {Donahue}, M., {et~al.} 2009, \apj, 693, 1142, \dodoi{10.1088/0004-637X/693/2/1142}

\bibitem[{{Tamura} {et~al.}(2019){Tamura}, {Fabian}, {Gandhi}, {Gu}, {Kamada}, {Kitayama}, {Loewenstein}, {Maeda}, {Matsushita}, {McCammon}, {Mitsuda}, {Nakashima}, {Porter}, {Pinto}, {Sato}, {Tombesi}, \& {Yamasaki}}]{Tamura19}
{Tamura}, T., {Fabian}, A.~C., {Gandhi}, P., {et~al.} 2019, \pasj, 71, 50, \dodoi{10.1093/pasj/psz023}

\bibitem[{{Tashiro} {et~al.}(2020){Tashiro}, {Maejima}, {Toda}, {Kelley}, {Reichenthal}, {Hartz}, {Petre}, {Williams}, {Guainazzi}, {Costantini}, {Fujimoto}, {Hayashida}, {Henegar-Leon}, {Holland}, {Ishisaki}, {Kilbourne}, {Loewenstein}, {Matsushita}, {Mori}, {Okajima}, {Porter}, {Sneiderman}, {Takei}, {Terada}, {Tomida}, {Yamaguchi}, {Watanabe}, {Akamatsu}, {Arai}, {Audard}, {Awaki}, {Babyk}, {Bamba}, {Bando}, {Behar}, {Bialas}, {Boissay-Malaquin}, {Brenneman}, {Brown}, {Canavan}, {Chiao}, {Comber}, {Corrales}, {Cumbee}, {de Vries}, {den Herder}, {Dercksen}, {Diaz-Trigo}, {DiPirro}, {Done}, {Dotani}, {Ebisawa}, {Eckart}, {Eckert}, {Eguchi}, {Enoto}, {Ezoe}, {Ferrigno}, {Fujita}, {Fukazawa}, {Furuzawa}, {Gallo}, {Gorter}, {Grim}, {Gu}, {Hagino}, {Hamaguchi}, {Hatsukade}, {Hawthorn}, {Hayashi}, {Hell}, {Hiraga}, {Hodges-Kluck}, {Horiuchi}, {Hornschemeier}, {Hoshino}, {Ichinohe}, {Iga}, {Iizuka}, {Ishida}, {Ishihama}, {Ishikawa}, {Ishimura}, {Jaffe}, {Kaastra}, {Kallman}, {Kara}, {Katsuda}, {Kenyon}, {Kimball},
  {Kitaguchi}, {Kitamoto}, {Kobayashi}, {Kobayashi}, {Kohmura}, {Kubota}, {Leutenegger}, {Li}, {Lockard}, {Maeda}, {Markevitch}, {Martz}, {Matsumoto}, {Matsuzaki}, {McCammon}, {McLaughlin}, {McNamara}, {Miko}, {Miller}, {Miller}, {Minesugi}, {Mitani}, {Mitsuishi}, {Mizumoto}, {Mizuno}, {Mukai}, {Murakami}, {Mushotzky}, {Nakajima}, {Nakamura}, {Nakazawa}, {Natsukari}, {Nigo}, {Nishioka}, {Nobukawa}, {Nobukawa}, {Noda}, {Odaka}, {Ogawa}, {Ohashi}, {Ohno}, {Ohta}, {Okamoto}, {Ota}, {Ozaki}, {Paltani}, {Plucinsky}, {Pottschmidt}, {Sampson}, {Sasaki}, {Sato}, {Sato}, {Sato}, {Sawada}, {Seta}, {Shibano}, {Shida}, {Shidatsu}, {Shigeto}, {Shinozaki}, {Shirron}, {Simionescu}, {Smith}, {Someya}, {Soong}, {Sugawara}, {Sugawara}, {Szymkowiak}, {Takahashi}, {Takeshima}, {Tamagawa}, {Tamura}, {Tanaka}, {Tanimoto}, {Terashima}, {Tsuboi}, {Tsujimoto}, {Tsunemi}, {Tsuru}, {Uchida}, {Uchida}, {Uchiyama}, {Ueda}, {Uno}, {Vink}, {Watanabe}, {Witthoeft}, {Wolfs}, {Yamada}, {Yamaoka}, {Yamasaki}, {Yamauchi}, {Yamauchi},
  {Yanagase}, {Yaqoob}, {Yasuda}, {Yoshida}, {Yoshioka}, \& {Zhuravleva}}]{Tashiro2020_XRISM}
{Tashiro}, M., {Maejima}, H., {Toda}, K., {et~al.} 2020, in Society of Photo-Optical Instrumentation Engineers (SPIE) Conference Series, Vol. 11444, Space Telescopes and Instrumentation 2020: Ultraviolet to Gamma Ray, ed. J.-W.~A. {den Herder}, S.~{Nikzad}, \& K.~{Nakazawa}, 1144422, \dodoi{10.1117/12.2565812}

\bibitem[{{Urban} {et~al.}(2015){Urban}, {Werner}, {Allen}, {Simionescu}, {Kaastra}, \& {Strigari}}]{2015MNRAS.451.2447U}
{Urban}, O., {Werner}, N., {Allen}, S.~W., {et~al.} 2015, \mnras, 451, 2447, \dodoi{10.1093/mnras/stv1142}

\bibitem[{{Walker} {et~al.}(2012){Walker}, {Fabian}, {Sanders}, \& {George}}]{2012MNRAS.424.1826W}
{Walker}, S.~A., {Fabian}, A.~C., {Sanders}, J.~S., \& {George}, M.~R. 2012, \mnras, 424, 1826, \dodoi{10.1111/j.1365-2966.2012.21282.x}

\bibitem[{{Walker} {et~al.}(2013){Walker}, {Fabian}, {Sanders}, {Simionescu}, \& {Tawara}}]{2013MNRAS.432..554W}
{Walker}, S.~A., {Fabian}, A.~C., {Sanders}, J.~S., {Simionescu}, A., \& {Tawara}, Y. 2013, \mnras, 432, 554, \dodoi{10.1093/mnras/stt497}

\bibitem[{{XRISM Collaboration} {et~al.}(2025{\natexlab{a}}){XRISM Collaboration}, {Audard}, {Awaki}, {Ballhausen}, {Bamba}, {Behar}, {Boissay-Malaquin}, {Brenneman}, {Brown}, {Corrales}, {Costantini}, {Cumbee}, {Done}, {Dotani}, {Ebisawa}, {Eckart}, {Eckert}, {Enoto}, {Eguchi}, {Ezoe}, {Foster}, {Fujimoto}, {Fujita}, {Fukazawa}, {Fukushima}, {Furuzawa}, {Gallo}, {Garc{\'\i}a}, {Gu}, {Guainazzi}, {Hagino}, {Hamaguchi}, {Hatsukade}, {Hayashi}, {Hayashi}, {Hell}, {Hodges-Kluck}, {Hornschemeier}, {Ichinohe}, {Ishida}, {Ishikawa}, {Ishisaki}, {Kaastra}, {Kallman}, {Kara}, {Katsuda}, {Kanemaru}, {Kelley}, {Kilbourne}, {Kitamoto}, {Kobayashi}, {Kohmura}, {Kubota}, {Leutenegger}, {Loewenstein}, {Maeda}, {Markevitch}, {Matsumoto}, {Matsushita}, {McCammon}, {McNamara}, {Mernier}, {Miller}, {Miller}, {Mitsuishi}, {Mizumoto}, {Mizuno}, {Mori}, {Mukai}, {Murakami}, {Mushotzky}, {Nakajima}, {Nakazawa}, {Ness}, {Nobukawa}, {Nobukawa}, {Noda}, {Odaka}, {Ogawa}, {Ogorzalek}, {Okajima}, {Ota}, {Paltani}, {Petre}, {Plucinsky},
  {Porter}, {Pottschmidt}, {Sato}, {Sato}, {Sawada}, {Seta}, {Shidatsu}, {Simionescu}, {Smith}, {Suzuki}, {Szymkowiak}, {Takahashi}, {Takeo}, {Tamagawa}, {Tamura}, {Tanaka}, {Tanimoto}, {Tashiro}, {Terada}, {Terashima}, {Trigo}, {Tsuboi}, {Tsujimoto}, {Tsunemi}, {Tsuru}, {Uchida}, {Uchida}, {Uchida}, {Uchiyama}, {Ueda}, {Uno}, {Vink}, {Watanabe}, {Williams}, {Yamada}, {Yamada}, {Yamaguchi}, {Yamaoka}, {Yamasaki}, {Yamauchi}, {Yamauchi}, {Yaqoob}, {Yoneyama}, {Yoshida}, {Yukita}, {Zhuravleva}, {Kondo}, {Werner}, {Pl{\v{s}}ek}, {Sun}, {Hosogi}, \& {Majumder}}]{XRISMCenI}
{XRISM Collaboration}, {Audard}, M., {Awaki}, H., {et~al.} 2025{\natexlab{a}}, \nat, 638, 365, \dodoi{10.1038/s41586-024-08561-z}

\bibitem[{{XRISM Collaboration} {et~al.}(2025{\natexlab{b}}){XRISM Collaboration}, {Audard}, {Awaki}, {Ballhausen}, {Bamba}, {Behar}, {Boissay-Malaquin}, {Brenneman}, {Brown}, {Corrales}, {Costantini}, {Cumbee}, {Diaz Trigo}, {Done}, {Dotani}, {Ebisawa}, {Eckart}, {Eckert}, {Eguchi}, {Enoto}, {Ezoe}, {Foster}, {Fujimoto}, {Fujita}, {Fukazawa}, {Fukushima}, {Furuzawa}, {Gallo}, {Garc{\'\i}a}, {Gu}, {Guainazzi}, {Hagino}, {Hamaguchi}, {Hatsukade}, {Hayashi}, {Hayashi}, {Hell}, {Hodges-Kluck}, {Hornschemeier}, {Ichinohe}, {Ishida}, {Ishikawa}, {Ishisaki}, {Kaastra}, {Kallman}, {Kara}, {Katsuda}, {Kanemaru}, {Kelley}, {Kilbourne}, {Kitamoto}, {Kobayashi}, {Kohmura}, {Kubota}, {Leutenegger}, {Loewenstein}, {Maeda}, {Markevitch}, {Matsumoto}, {Matsushita}, {McCammon}, {McNamara}, {Mernier}, {Miller}, {Miller}, {Mitsuishi}, {Mizumoto}, {Mizuno}, {Mori}, {Mukai}, {Murakami}, {Mushotzky}, {Nakajima}, {Nakazawa}, {Ness}, {Nobukawa}, {Nobukawa}, {Noda}, {Odaka}, {Ogawa}, {Ogorzalek}, {Okajima}, {Ota}, {Paltani},
  {Petre}, {Plucinsky}, {Porter}, {Pottschmidt}, {Sato}, {Sato}, {Sawada}, {Seta}, {Shidatsu}, {Simionescu}, {Smith}, {Suzuki}, {Szymkowiak}, {Takahashi}, {Takeo}, {Tamagawa}, {Tamura}, {Tanaka}, {Tanimoto}, {Tashiro}, {Terada}, {Terashima}, {Tsuboi}, {Tsujimoto}, {Tsunemi}, {Tsuru}, {Uchida}, {Uchida}, {Uchida}, {Uchiyama}, {Ueda}, {Uno}, {Vink}, {Watanabe}, {Williams}, {Yamada}, {Yamada}, {Yamaguchi}, {Yamaoka}, {Yamasaki}, {Yamauchi}, {Yamauchi}, {Yaqoob}, {Yoneyama}, {Yoshida}, {Yukita}, {Zhuravleva}, {Bartalesi}, {Ettori}, {Kosarzycki}, {Lovisari}, {Rose}, {Sarkar}, {Sun}, \& {Tamhane}}]{XRISM2029a}
---. 2025{\natexlab{b}}, \apjl, 982, L5, \dodoi{10.3847/2041-8213/ada7cd}

\bibitem[{{XRISM Collaboration} {et~al.}(2025{\natexlab{c}}){XRISM Collaboration}, {Audard}, {Awaki}, {Ballhausen}, {Bamba}, {Behar}, {Boissay-Malaquin}, {Brenneman}, {Brown}, {Corrales}, {Costantini}, {Cumbee}, {Diaz Trigo}, {Done}, {Dotani}, {Ebisawa}, {Eckart}, {Eckert}, {Eguchi}, {Enoto}, {Ezoe}, {Foster}, {Fujimoto}, {Fujita}, {Fukazawa}, {Fukushima}, {Furuzawa}, {Gallo}, {Garc{\'\i}a}, {Gu}, {Guainazzi}, {Hagino}, {Hamaguchi}, {Hatsukade}, {Hayashi}, {Hayashi}, {Hell}, {Hodges-Kluck}, {Hornschemeier}, {Ichinohe}, {Ishi}, {Ishida}, {Ishikawa}, {Ishisaki}, {Kaastra}, {Kallman}, {Kara}, {Katsuda}, {Kanemaru}, {Kelley}, {Kilbourne}, {Kitamoto}, {Kobayashi}, {Kohmura}, {Kubota}, {Leutenegger}, {Loewenstein}, {Maeda}, {Markevitch}, {Matsumoto}, {Matsushita}, {McCammon}, {McNamara}, {Mernier}, {Miller}, {Miller}, {Mitsuishi}, {Mizumoto}, {Mizuno}, {Mori}, {Mukai}, {Murakami}, {Mushotzky}, {Nakajima}, {Nakazawa}, {Ness}, {Nobukawa}, {Nobukawa}, {Noda}, {Odaka}, {Ogawa}, {Ogorzalek}, {Okajima}, {Ota}, {Paltani},
  {Petre}, {Plucinsky}, {Porter}, {Pottschmidt}, {Sato}, {Sato}, {Sawada}, {Seta}, {Shidatsu}, {Simionescu}, {Smith}, {Suzuki}, {Szymkowiak}, {Takahashi}, {Takeo}, {Tamagawa}, {Tamura}, {Tanaka}, {Tanimoto}, {Tashiro}, {Terada}, {Terashima}, {Tsuboi}, {Tsujimoto}, {Tsunemi}, {Tsuru}, {Uchida}, {Uchida}, {Uchida}, {Uchiyama}, {Ueda}, {Uno}, {Vink}, {Watanabe}, {Williams}, {Yamada}, {Yamada}, {Yamaguchi}, {Yamaoka}, {Yamasaki}, {Yamauchi}, {Yamauchi}, {Yaqoob}, {Yoneyama}, {Yoshida}, {Yukita}, {Zhuravleva}, {Bartalesi}, {Ettori}, {Kosarzycki}, {Lovisari}, {Rose}, {Sarkar}, {Sun}, \& {Tamhane}}]{XRISM2029b}
---. 2025{\natexlab{c}}, arXiv e-prints, arXiv:2505.06533, \dodoi{10.48550/arXiv.2505.06533}

\bibitem[{{XRISM Collaboration} {et~al.}(2025{\natexlab{d}}){XRISM Collaboration}, {Audard}, {Awaki}, {Ballhausen}, {Bamba}, {Behar}, {Boissay-Malaquin}, {Brenneman}, {Brown}, {Corrales}, {Costantini}, {Cumbee}, {Diaz Trigo}, {Done}, {Dotani}, {Ebisawa}, {Eckart}, {Eckert}, {Eguchi}, {Enoto}, {Ezoe}, {Foster}, {Fujimoto}, {Fujita}, {Fukazawa}, {Fukushima}, {Furuzawa}, {Gallo}, {Garc{\'\i}a}, {Gu}, {Guainazzi}, {Hagino}, {Hamaguchi}, {Hatsukade}, {Hayashi}, {Hayashi}, {Hell}, {Hodges-Kluck}, {Hornschemeier}, {Ichinohe}, {Ishi}, {Ishida}, {Ishikawa}, {Ishisaki}, {Kaastra}, {Kallman}, {Kara}, {Katsuda}, {Kanemaru}, {Kelley}, {Kilbourne}, {Kitamoto}, {Kobayashi}, {Kohmura}, {Kubota}, {Leutenegger}, {Loewenstein}, {Maeda}, {Markevitch}, {Matsumoto}, {Matsushita}, {McCammon}, {McNamara}, {Mernier}, {Miller}, {Miller}, {Mitsuishi}, {Mizumoto}, {Mizuno}, {Mori}, {Mukai}, {Murakami}, {Mushotzky}, {Nakajima}, {Nakazawa}, {Ness}, {Nobukawa}, {Nobukawa}, {Noda}, {Odaka}, {Ogawa}, {Ogorza{\l}ek}, {Okajima}, {Ota},
  {Paltani}, {Petre}, {Plucinsky}, {Porter}, {Pottschmidt}, {Sato}, {Sato}, {Sawada}, {Seta}, {Shidatsu}, {Simionescu}, {Smith}, {Suzuki}, {Szymkowiak}, {Takahashi}, {Takeo}, {Tamagawa}, {Tamura}, {Tanaka}, {Tanimoto}, {Tashiro}, {Terada}, {Terashima}, {Tsuboi}, {Tsujimoto}, {Tsunemi}, {Tsuru}, {T{\"u}mer}, {Uchida}, {Uchida}, {Uchida}, {Uchiyama}, {Ueda}, {Ueda}, {Uno}, {Vink}, {Watanabe}, {Williams}, {Yamada}, {Yamada}, {Yamaguchi}, {Yamaoka}, {Yamasaki}, {Yamauchi}, {Yamauchi}, {Yaqoob}, {Yoneyama}, {Yoshida}, {Yukita}, {Zhuravleva}, {Fabian}, {Nelson}, {Okabe}, {Pillepich}, {Potter}, {Regamey}, {Sakai}, {Shishido}, {Truong}, {Wik}, \& {Zuhone}}]{XRISMcomaa}
---. 2025{\natexlab{d}}, \apjl, 985, L20, \dodoi{10.3847/2041-8213/add2f6}

\bibitem[{{XRISM Collaboration} {et~al.}(2025{\natexlab{e}}){XRISM Collaboration}, {Audard}, {Awaki}, {Ballhausen}, {Bamba}, {Behar}, {Boissay-malaquin}, {Brenneman}, {Brown}, {Corrales}, {Costantini}, {Cumbee}, {Diaz Trigo}, {Done}, {Dotani}, {Ebisawa}, {Eckart}, {Eckert}, {Eguchi}, {Enoto}, {Ezoe}, {Foster}, {Fujimoto}, {Fujita}, {Fukazawa}, {Fukushima}, {Furuzawa}, {Gallo}, {Garc{\'\i}a}, {Gu}, {Guainazzi}, {Hagino}, {Hamaguchi}, {Hatsukade}, {Hayashi}, {Hayashi}, {Hell}, {Hodges-kluck}, {Hornschemeier}, {Ichinohe}, {Ishi}, {Ishida}, {Ishikawa}, {Ishisaki}, {Kaastra}, {Kallman}, {Kara}, {Katsuda}, {Kanemaru}, {Kelley}, {Kilbourne}, {Kitamoto}, {Kobayashi}, {Kohmura}, {Kubota}, {Leutenegger}, {Loewenstein}, {Maeda}, {Markevitch}, {Matsumoto}, {Matsushita}, {Mccammon}, {Mcnamara}, {Mernier}, {Miller}, {Miller}, {Mitsuishi}, {Mizumoto}, {Mizuno}, {Mori}, {Mukai}, {Murakami}, {Mushotzky}, {Nakajima}, {Nakazawa}, {Ness}, {Nobukawa}, {Nobukawa}, {Noda}, {Odaka}, {Ogawa}, {Ogorzalek}, {Okajima}, {Ota}, {Paltani},
  {Petre}, {Plucinsky}, {Porter}, {Pottschmidt}, {Sato}, {Sato}, {Sawada}, {Seta}, {Shidatsu}, {Simionescu}, {Smith}, {Suzuki}, {Szymkowiak}, {Takahashi}, {Takeo}, {Tamagawa}, {Tamura}, {Tanaka}, {Tanimoto}, {Tashiro}, {Terada}, {Terashima}, {Tsuboi}, {Tsujimoto}, {Tsunemi}, {Tsuru}, {Uchida}, {Uchida}, {Uchida}, {Uchiyama}, {Ueda}, {Uno}, {Vink}, {Watanabe}, {Williams}, {Yamada}, {Yamada}, {Yamaguchi}, {Yamaoka}, {Yamasaki}, {Yamauchi}, {Yamauchi}, {Yaqoob}, {Yoneyama}, {Yoshida}, {Yukita}, {Zhuravleva}, {Seppi}, {Aihara}, \& {Omiya}}]{XRISMA2319}
---. 2025{\natexlab{e}}, arXiv e-prints, arXiv:2508.05067, \dodoi{10.48550/arXiv.2508.05067}

\bibitem[{{XRISM Collaboration} {et~al.}(2025{\natexlab{f}}){XRISM Collaboration}, {Audard}, {Awaki}, {Ballhausen}, {Bamba}, {Behar}, {Boissay-Malaquin}, {Brenneman}, {Brown}, {Corrales}, {Costantini}, {Cumbee}, {Diaz Trigo}, {Done}, {Dotani}, {Ebisawa}, {Eckart}, {Eckert}, {Eguchi}, {Enoto}, {Ezoe}, {Foster}, {Fujimoto}, {Fujita}, {Fukazawa}, {Fukushima}, {Furuzawa}, {Gallo}, {Garcia}, {Gu}, {Guainazzi}, {Hagino}, {Hamaguchi}, {Hatsukade}, {Hayashi}, {Hayashi}, {Hell}, {Hodges-Kluck}, {Hornschemeier}, {Ichinohe}, {Ishi}, {Ishida}, {Ishikawa}, {Ishisaki}, {Kaastra}, {Kallman}, {Kara}, {Katsuda}, {Kanemaru}, {Kelley}, {Kilbourne}, {Kitamoto}, {Kobayashi}, {Kohmura}, {Kubota}, {Leutenegger}, {Loewenstein}, {Maeda}, {Markevitch}, {Matsumoto}, {Matsushita}, {McCammon}, {McNamara}, {Mernier}, {Miller}, {Miller}, {Mitsuishi}, {Mizumoto}, {Mizuno}, {Mori}, {Mukai}, {Murakami}, {Mushotzky}, {Nakajima}, {Nakazawa}, {Ness}, {Nobukawa}, {Nobukawa}, {Noda}, {Odaka}, {Ogawa}, {Ogorzalek}, {Okajima}, {Ota}, {Paltani},
  {Petre}, {Plucinsky}, {Porter}, {Pottschmidt}, {Sato}, {Sato}, {Sawada}, {Seta}, {Shidatsu}, {Simionescu}, {Smith}, {Suzuki}, {Szymkowiak}, {Takahashi}, {Takeo}, {Tamagawa}, {Tamura}, {Tanaka}, {Tanimoto}, {Tashiro}, {Terada}, {Terashima}, {Tsuboi}, {Tsujimoto}, {Tsunemi}, {Tsuru}, {Tumer}, {Uchida}, {Uchida}, {Uchida}, {Uchiyama}, {Ueda}, {Uno}, {Vink}, {Watanabe}, {Williams}, {Yamada}, {Yamada}, {Yamaguchi}, {Yamaoka}, {Yamasaki}, {Yamauchi}, {Yamauchi}, {Yaqoob}, {Yoneyama}, {Yoshida}, {Yukita}, {Zhuravleva}, {Bellomi}, {Drury}, {Heinrich}, {Hlavacek-Larrondo}, {Meunier}, {Migkas}, {Shefler}, {Stancil}, {Truong}, {Ueda}, {Vigneron}, {Zhang}, \& {ZuHone}}]{XRISMPers}
---. 2025{\natexlab{f}}, arXiv e-prints, arXiv:2509.04421, \dodoi{10.48550/arXiv.2509.04421}

\bibitem[{{Yin} {et~al.}(2025){Yin}, {Fujita}, {Ezoe}, \& {Ishisaki}}]{2025arXiv250304726Y}
{Yin}, W., {Fujita}, Y., {Ezoe}, Y., \& {Ishisaki}, Y. 2025, arXiv e-prints, arXiv:2503.04726, \dodoi{10.48550/arXiv.2503.04726}

\bibitem[{{Zhang} {et~al.}(2025){Zhang}, {Zhuravleva}, {Heinrich}, {Bellomi}, {Truong}, {ZuHone}, {Churazov}, {Eckart}, {Fujita}, {Hlavacek-Larrondo}, {Ichinohe}, {Markevitch}, {Matsushita}, {Mernier}, {Miller}, {Mori}, {Nakajima}, {Ogorzalek}, {Porter}, {T{\"u}mer}, {Ueda}, \& {Werner}}]{2025arXiv251012782Z}
{Zhang}, C., {Zhuravleva}, I., {Heinrich}, A., {et~al.} 2025, arXiv e-prints, arXiv:2510.12782.
\newblock \doarXiv{2510.12782}

\bibitem[{{Zhou} {et~al.}(2024){Zhou}, {Takhistov}, \& {Mitsuda}}]{2024ApJ...976..238Z}
{Zhou}, Y., {Takhistov}, V., \& {Mitsuda}, K. 2024, \apj, 976, 238, \dodoi{10.3847/1538-4357/ad83cf}

\end{thebibliography}
\bibliographystyle{aasjournal}

\end{document}